\documentclass[12pt]{article}
\usepackage{amssymb,amscd,array}
\catcode `\@=11
\@addtoreset{equation}{section}

\def\harr#1#2{\smash{\mathop{\hbox to .5in{\rightarrowfill}}
\limits^{\scriptstyle#1}_{\scriptstyle#2}}}

\def\harrl#1#2{\smash{\mathop{\hbox to .5in{\leftarrowfill}}
\limits^{\scriptstyle#1}_{\scriptstyle#2}}}

\def\qed{\hspace*{\fill}\rule{3mm}{3mm}}

\newcommand{\be}{\begin{equation}}
\newcommand{\ee}{\end{equation}}
\newcommand{\bea}{\begin{eqnarray}}
\newcommand{\eea}{\end{eqnarray}}
\def\d{\partial}

\newcommand{\R}{\mathbb{R}}

\newcommand{\C}{\mathbb{C}}
\newtheorem{thm}{Theorem}[section]
\newtheorem{rem}[thm]{Remark}
\newtheorem{lemma}[thm]{Lemma}
\newtheorem{cor}[thm]{Corollary}
\newtheorem{prop}[thm]{Proposition}
\begin{document}
\begin{titlepage}
\begin{center}
{\bf \Large{A Supersymmetric Extension of Quantum Gauge Theory\\}}
\end{center}
\vskip 1.0truecm
\centerline{Dan Radu Grigore
\footnote{e-mail: grigore@physik.unizh.ch}
\footnote{Permanent address: Dept. Theor. Phys., Inst. Atomic Phys.,
Bucharest-M\u agurele, MG 6, Rom\^ania}
and G\"unter Scharf
\footnote{e-mail: scharf@physik.unizh.ch}}
\vskip5mm
\centerline{Institute of Theor. Phys., University of Z\"urich}
\centerline{Winterthurerstr., 190, Z\"urich CH-8057, Switzerland}
\vskip 2cm
\bigskip \nopagebreak
\begin{abstract}
\noindent
We consider a supersymmetric extension of quantum gauge theory based on
a 
 vector multiplet containing supersymmetric partners of spin $3/2$
for
 the vector fields. The constructions of the model follows closely
the usual construction of gauge models in the Epstein-Glaser framework
for perturbative field theory. Accordingly, all the arguments are
completely of quantum nature without reference to a classical
supersymmetric
 theory.
 
As an application we consider the
supersymmetric electroweak theory. The resulting self-couplings of
the gauge bosons agree with the standard model up to a divergence.
\end{abstract}

\newpage\setcounter{page}1
\end{titlepage}
\section{Introduction}

The supersymmetric gauge theories are usually constructed using the so-called
vector supersymmetric multiplet \cite{WB}, \cite{Wes} \cite{Wei}, \cite{Bi},
\cite{Fi}, \cite{Li}, \cite{So}, \cite{Pi1}, \cite{Pi2}, \cite{Sa}, 
 etc. In
fact, this is not the only logical possibility. 
 If one wants to obtain
a supersymmetric theory such that vector fields 
 (describing the usual
gauge fields) appear, then one has to include the usual 
 vector field
into a supersymmetric multiplet. As noticed in \cite{CS}, 
\cite{CGS}, the
analysis of the unitary irreducible representations of the
$N = 1$
supersymmetric extension of the Poincar\'e group gives two irreducible
massive representations
$\Omega_{1/2}$
and
$\Omega_{1}$
containing a spin 1 system. (See \cite{RY1}, \cite{RY2} for
 a clear derivation
of the SUSY IRREPS). The standard vector multiplet is 
constructed such that the
one-particle subspace of the Fock space carries the representation
$\Omega_{1/2}$.
Surprisingly enough, it is hard to build a corresponding fully consistent
theory with all the usual properties. The other possibility is to
 construct
a supersymmetric multiplet for which the associated Fock space has
$\Omega_{1}$
as the one-particle subspace of the Fock space. The construction in this case
is natural and straightforward. In \cite{CS}, \cite{CGS} such a
multiplet was constructed; the content of this multiplet was a spin 1/2, a
(complex) spin
 1 and a spin 3/2 fields. In this paper we will use a multiplet
containing only
 a complex vector and a Rarita-Schwinger field (without the
transversality conditions) which is in fact related to the previous one. We
will prove that such a multiplet can be the
 basis for a supersymmetric
extension of quantum gauge theory. In fact, this
 multiplet is distinguished by
the property that its gauge structure involving
 ghosts, anti-ghosts and
unphysical scalar (Goldstone) fields is precisely the
 same as in ordinary gauge
theory.

Our supermultiplet of free fields in which the spin $3/2$ field is the
supersymmetric partner of a complex vector fields is perfectly well defined
from the mathematical point of view. This multiplet is distinct from another
multiplet appearing in the literature and contaning a spin $3/2$ field, namely
the supergravity multiplet.

We will do the analysis entirely in the quantum framework avoiding the usual
approach based on  quantizing a classical supersymmetric  theory.  In this way
we avoid completely the usual complications associated  to the proper
mathematical definition of a super-manifold and the quantization procedures
\cite{Fr}, \cite{IAS}. This point of view is rather new in the  literature
\cite{Lo}, \cite{O}, \cite{NK}, \cite{CGS}, \cite{Gr2}.  The construction of
the $S$-matrix will be done in the spirit of Epstein-Glaser construction
\cite{Sc}.

Let us outline the mathematical framework used in this paper. The description
of higher spin fields will be done in the indefinite metric approach
(Gupta-Bleuler). That is, we construct a Hilbert space
${\cal H}$
with a non-degenerate 
sesqui-linear form
$<\cdot, \cdot>$
and a gauge charge operator $Q$ verifying
$Q^{2} = 0$;
the form
$<\cdot, \cdot>$
becomes positively defined when restricted to a factor Hilbert space
${\cal H}_{phys} \equiv Ker(Q)/Im(Q)$
which will be the physical space of our problem.
The interaction Lagrangian
$T(x)$
will be some Wick polynomial acting in the total Hilbert space
${\cal H}$
and verifying the conditions
\bea
[Q, T(x)]= i\d_\mu T^{\mu}(x)
\label{gauge}
\eea
for some Wick polynomials
$T^{\mu}(x)
$;
this condition guarantees that the interaction Lagrangian
$T(x)$
factorizes to the physical Hilbert space
$Ker(Q)/Im(Q)$
in the adiabatic limit, i.e. after integration over $x$. The condition
(\ref{gauge}) is equivalent to the usual condition of current conservation
if one considers for instance the coupling of the gauge field to
an external current of matter fields. ( A natural generalization of
(\ref{gauge}) can be provided for a supersymmetric theory.) We remark in this
context that there are some arguments in the literature about the impossibility
to construct consistent quantum field theories for higher spins
($s \geq 5/2$).
These arguments are based on the impossibility to construct the corresponding
conserved current \cite{Wei0}. In our formalism this will mean that although
the construction of the Hilbert space
${\cal H}_{phys} \equiv Ker(Q)/Im(Q)$
is still possible for higher spins, one will not be able to find interaction
Lagrangians which are solutions of the equation (\ref{gauge}). This conjecture
is worthwile investigating. Moreover, it is argued that a consistent theory for
spin $3/2$ can be built only in a supersymmetric context and such that the spin
$3/2$ is the supersymmetric partner of the graviton \cite{Pe}.
We can construct toy models (for instance a supersymmetric generalization of
the Abelian Higgs model) in which the spin $3/2$ is coupled in a supersymmetric
non-trivial way to the complex vector partner, but it is less clear if the same
is possible for a realistic model generalizing the electroweak theory.
In Section \ref{ew} of our paper we avoid this no-go result, brought to our
attention by the referee, by breaking the supersymmetry in the coupling
(interaction Lagrangian). This is necessary anyway in a realistic theory
because no known particle has a superpartner of equal mass.

In Section \ref{qsr}  we give a general discussion about supersymmetric
multiplets and
 the associated superfields and provide an elementary
derivation of the
 supersymmetric Ward identities. The construction of
the superfield associated to a given supersymmetric multiplet is not an
unique operation; we make a choice which is more convenient from the
point of view of renormalization theory. In Section \ref{multiplets}  we
describe the multiplets used in
 the construction of the gauge theory
and determine the action of the gauge charge. We
 use the point of view
of \cite{LW}. In Section \ref{ew} we discuss
 the supersymmetric
extension of electroweak theory. One of the main virtues of
 our model
is that supersymmetry of the free aymptotic fields and gauge invariance
fix the
 interaction Lagrangian quite drastically: the number of free parameters
is
 essentially the same as for the usual electro-weak model. This is in
contrast
 with the usual approaches to supersymmetric extensions of the
standard model
 for which the number of parameters increases
dramatically.

\section{Quantum Supersymetric Theory \label{qsr}}

\subsection{Supersymetric Multiplets and Superfields\label{susy-multiplets}}

Let us define from the very beginning what we mean by a supersymmetric
theory in a pure quantum context. We will not consider extended
supersymmetries here.

As a matter of convention, in the following we raise and lower Minkowski
indices with the Minkowski pseudo-metric
$g_{\mu\nu} = g^{\mu\nu}$
with diagonal
$1,-1,-1,-1$;
we also  raise and lower Weyl indices with the anti-symmetric
$SL(2,\C)$-invariant
tensor
$\epsilon_{ab} = - \epsilon^{ab}; \quad \epsilon_{12} = 1$
and we use summation over dummy indices. By
$
SL(2,\C) \ni A \mapsto \delta(A) \in {\cal L}^{\uparrow}_{+}
$
we denote the universal covering homomorphism of the proper orthochronous
Lorentz group.

Suppose that we have a Wightman theory
$({\cal H}, (\cdot, \cdot), U_{a,A}, \Omega, b_{j}, f_{A}), \quad
j = 1,\dots,N_{B}, \quad A = 1,\dots,N_{F}$
where
${\cal H}$
is a Hilbert space with the scalar product
$(\cdot, \cdot)$,
$U_{a,A}$ is a unitary irreducible representation of
$inSL(2,\C)$
the universal covering group of the proper orthochronous Poincar\'e group such
that
$a \in \R^{4}$
is translation in the Minkowski space and
$A \in SL(2,\C)$,
$\Omega$
is the vacuum and
$b_{j}$ (resp. $f_{A}$)
are the quantum bosonic (resp. fermionic) fields. It is natural to assume
that the fields are linearly independent (over the ring of partial derivative
operators), that is only equations of motion pertaining to a single field
are allowed. The transformation Lorentz properties of these fields 
are
encoded in two finite dimensional representations
$D_{B}(A)$
and
$D_{F}(A)$
of dimension
$N_{B}$
and
$N_{F}$
respectively.

Sometimes it is necessary to extend somewhat this framework: one considers in
${\cal H}$
besides the usual positive definite scalar product a non-degenerate
sesqui-linear form
$<\cdot, \cdot>$
which becomes positively defined when restricted to a factor Hilbert space
$Ker(Q)/Im(Q)$
where $Q$ is some {\it gauge charge}. We denote with
$A^{\dagger}$
the adjoint of the operator $A$ with respect to
$<\cdot, \cdot>$.
As a matter of convenience one can assume, without
losing generality, that the bosonic fields are Hermitian and all the
fermionic fields are Majorana:
\be
(b_{j})^{\dagger} = b_{j}, \quad j = 1,\dots,N_{B} \qquad
(f_{Aa_{1},\dots,a_{r}})^{\dagger} = \bar{f}_{A\bar{a_{1}},\dots,\bar{a}_{r}},
\quad A = 1,\dots,N_{F}, \quad a=1,2
\ee
where we use Weyl notations for the Fermi fields.

Suppose that in the Hilbert space
${\cal H}$
we also have the operators
$Q_{a}, \quad a = 1,2$
such that:

(i) the following relations are verified:
\be
Q_{a} \Omega = 0, \quad \bar{Q}_{\bar{a}} \Omega = 0
\label{vac}
\ee
and
\bea
\{ Q_{a} , Q_{b} \} = 0, \quad
\{ Q_{a} , \bar{Q}_{\bar{b}} \} = 2 \sigma^{\mu}_{a\bar{b}} P_{\mu}, \quad
[ Q_{a}, P_{\mu} ] = 0, \quad
U_{a,A}^{-1} Q_{b} U_{a,A} = {A_{b}}^{c} Q_{c}.
\label{SUSY}
\eea

Here
$P_{\mu} \quad = - i~\partial_{\mu}$
are the infinitesimal generators of the translation group,
$\sigma^{\mu}$
are the usual Pauli matrices and
\be
\bar{Q}_{\bar{b}} \equiv (Q_{b})^{\dagger}.
\ee

(ii) The following commutation relations are true:
\bea
i [ Q_{a}, b_{k} ] = \sum_{A} p_{a;kA}(\partial) f_{A},
\qquad
\{ Q_{a}, f_{A} \} = \sum_{j} q_{a;Aj}(\partial) b_{j}
\label{tensor}
\eea
where
$p_{a}$
and
$q_{a}$
are matrix-valued polynomials in the partial derivatives (with constant
coefficients). These relations express the tensor properties of the fields
with respect to (infinitesimal) supersymmetry transformations.

If this conditions are true we say that
$Q_{a}$
are {\it super-charges} and
$b_{j}, f_{A}$
are forming a {\it supersymmetric multiplet}. A natural notion of
{\it irreducibility} can be defined for any supersymmetric multiplet.
There are no general classification results for the supersymmetric
multiplets even in the case when we are dealing with free fields. One can
obtain however on general grounds relations between the numbers
$N_{B}$
and
$N_{F}$
expressing the well-known folklore about the equality of Bosonic
and Fermionic degrees of freedom (see for instance \cite{Wei} Section 26.2.)

However, it can be already said that the matrix-valued operators
$p_{a}$
and
$q_{a}$
are subject to various constraints. Let us describe them.
\begin{itemize}
\item
An immediate consistency condition follows from the compatibility of
(\ref{tensor}) with Lorentz transformations: we get that these polynomials
should be Lorentz covariant i.e. for all
$A \in SL(2,\C)$
we should have:
\be
p_{a}(\delta(A)\cdot\partial)
= {A_{a}}^{b}~D_{B}(A) \otimes D_{F}(A)~p_{b}(\partial), \qquad
q_{a}(\delta(A)\cdot\partial)
= {A_{a}}^{b}~D_{F}(A) \otimes D_{B}(A)~q_{b}(\partial).
\label{lorentz}
\ee
\item
Next, we start from the fact that the Hilbert space of the model is
generated by vectors of the type
\bea
\Psi = \prod b_{j_{p}}(x_{p})~\prod f_{A_{q}a_{q}}(x_{q})~
\prod \bar{f}_{A_{r}\bar{a}_{r}}(x_{r}) \Omega \in {\cal H}.
\label{hilbert}
\eea
(this is in fact one of the Wightman axioms). The action of the supercharges
$Q_{a}, \quad \bar{Q}_{\bar{a}}$
is determined by (\ref{tensor}): one commutes the supercharge operators to
the right till they hit the vacuum and then one applies (\ref{vac}).
However, the supercharges are not independent: they are constrained by the
relations from (\ref{SUSY}) and we should check that we do not get
a contradiction. The consistency relations are given by the (graded) Jacobi
identities combined with (\ref{SUSY}) and the well-known relation:
\be
[P_{\mu}, b_{j} ] = - i~\partial_{\mu} b_{j}, \quad
[P_{\mu}, f_{A} ] = - i~\partial_{\mu} f_{A}.
\ee

As a result must we have:
\bea
\left\{ Q_{a} , [ Q_{b}, b_{j} ] \right\} = - ( a \leftrightarrow b)
\nonumber \\
\left[ Q_{a} , \{ Q_{b}, f_{A} \} \right] = - ( a \leftrightarrow b), \qquad
\left[ Q_{a} , \{ Q_{b}, \bar{f}_{A} \} \right] = - ( a \leftrightarrow b
)
\nonumber \\
\left\{ Q_{a} , [ \bar{Q}_{\bar{b}}, b_{j} ] \right\}
+ \{ \bar{Q}_{\bar{b}} , [ Q_{a}, b_{j} ] \} =
- 2i~\sigma^{\mu}_{a\bar{b}}~\partial_{\mu} b_{j},
\nonumber \\
\left[ Q_{a} , \{ \bar{Q}_{\bar{b}}, f_{A} \} \right]
+ [ \bar{Q}_{\bar{b}} , \{ Q_{a}, f_{A} \} ] =
- 2i~\sigma^{\mu}_{a\bar{b}}~\partial_{\mu} f_{A}.
\label{susy+lorentz}
\eea
\item
If the fields of the multiplet are free fields i.e. relations of the type
\bea
e_{j}(\partial) b_{j} = 0, \quad  j = 1,\dots, N_{B}
\nonumber \\
E_{A}(\partial) f_{A} = 0, \quad  A = 1,\dots, N_{F}
\label{motion}
\eea
are valid (where
$e_{j}, \quad j= 1,\dots,N_{B}$
and
$E_{A}, \quad A = 1,\dots,N_{F}$
are some linear partial differential operators, with constant coefficients,
depending on the masses
$m_{j}$
and
$M_{A}$
of the fields), then commuting (resp. anticommuting) with the supercharges and
using (\ref{tensor}) 
new constraints on the polynomials
$p_{a}$
and
$q_{a}$
show up namely the polynomial
$e_{j}~p_{a;jB}$
(resp.
$E_{A}~q_{a;Ak}$)
should have $E_{B}$ (resp. $e_{k}$) as a factor:
\be
e_{j}~p_{a;jB} = E_{B}~e', \qquad E_{A}~q_{a;Ak} = e_{k}~E'
\label{factor}
\ee
for some polynomials $e'$ and $E'$;
otherwise we would get new equations of motion for the free fields of the type
$
e_{j}(\partial)~p_{a;jB}(\partial) f_{B} = 0, \qquad
E_{A}(\partial)~q_{a;Ak}(\partial) b_{k} = 0.
$
\item
Also, for free fields causal (anti)commutations are valid \cite{U}
\bea
\left[ b_{j}(x), b_{k}(y) \right] = - i~d_{jk}(\partial) D(x-y),
\nonumber \\
\left\{ f_{A}(x), f_{B}(y) \right\} = - i~d_{AB}(\partial) D(x-y),
\nonumber \\
\left[ b_{j}(x), f_{A}(y) \right] = 0
\label{CCR}
\eea
where
$D$ is the Pauli-Jordan causal function and
$d_{jk}, \quad D_{AB}$
are polynomials in the partial differential operators (with constant
coefficients). We note that for a ghost multiplet the r\^ole of the
commutator and anti-commutator in the first two relations should be reversed.
The (anti)commutation relations have the implication that one and the same
vector from the Hilbert space
${\cal H}$
can be expressed in the form (\ref{hilbert}) in two distinct way. This means
that the supercharges are well defined {\it via} (\ref{SUSY}) {\it iff} some
new consistency relations are valid following again from graded Jacobi
identities; the non-trivial ones are of the form:
\bea
[ b_{j}(x), \{ f_{A}(y), Q_{a} \} ] = - \{ f_{A}(y), [ {Q}_{a}, b_{j}(x) ] \}
\label{susy+CCR}
\eea
\item
Finally, if a gauge supercharge $Q$ is present in the model, then it is usually
determined by relations of the type (\ref{tensor}) involving ghost fields also,
so it means that we must impose consistency relations of the same type as
above. Moreover, it is desirable to have
\be
\{ Q, Q_{a} \} = 0, \quad \{ Q, \bar{Q}_{\bar{a}} \} = 0.
\label{susy+gauge}
\ee
and this implies new consistency relations of the type (\ref{susy+lorentz})
with one of the supercharges replaced by the gauge charge:
\bea
~\left\{ Q_{a} , [ Q, b_{j} ] \right\} = - \left\{ Q , [ Q_{a}, b_{j} ] \right\},
\nonumber \\
~\left[ Q_{a} , \{ Q, f_{A} \} \right] = - \left[ Q, \{ Q_{a}, f_{A} \} \right],
\qquad
\left[ Q_{a} , \{ Q, \bar{f}_{A} \} \right]
= - \left[ Q, \{ Q_{a}, \bar{f}_{A} \} \right],
\label{susy+gauge1}
\eea
\end{itemize}
\begin{rem}
Let us note that all these conditions are of pure quantum nature i.e. they can
be understood only for a pure quantum model. It is not clear to us if the
usual approaches to supersymmetric model, based on the quantization of some
classical field theory (in which Fermi fields are modelled as odd degrees
of freedom on some super-manifold) guarantees automatically that these
consistency relations are true.
\end{rem}

It seems to be an essential point to describe supersymmetric theories in
{\it superspace} \cite{SS1}, \cite{SS2}. We do this in the following way.
We consider the space
${\cal H}_{G} \equiv {\cal G} \otimes {\cal H}$
where
${\cal G}$
is a Grassmann algebra generated by Weyl anticommuting spinors
$\theta_{a}$
and their complex conjugates
$\bar{\theta}_{\bar{a}} = (\theta_{a})^{*}$
and perform a Klein transform such that the Grassmann parameters
$\theta_{a}$
are anti-commuting with all fermionic fields, the supercharges and
the gauge charge. The operators acting in
${\cal H}_{G}$
are called {\it superfields}. Of special interest are the superfields
constructed as in \cite{CS}, \cite{CGS} according to the formul\ae:
\bea
B_{j}(x,\theta,\bar{\theta}) \equiv
W_{\theta,\bar{\theta}}~b_{j}(x)~W_{\theta,\bar{\theta}}^{-1},
\nonumber \\
F_{A}(x,\theta,\bar{\theta}) \equiv
W_{\theta,\bar{\theta}}~f_{A}(x)~W_{\theta,\bar{\theta}}^{-1},
\label{superfields}
\eea
where
\be
W_{\theta,\bar{\theta}} \equiv
\exp\left(i\theta^{a} Q_{a} - i\bar{\theta}^{\bar{a}} \bar{Q}_{\bar{a}}\right)
\label{W-expo}
\ee
and we interpret the exponential as a (finite) Taylor series.
It is a 
remarkable fact that only such type of superfields are really
necessary,
 so in the following, when referring to superfields we mean
expressions given
 by (\ref{superfields}). We will call them {\it super-Bose}
and respectively 
 {\it super-Fermi} fields.
 For convenience we will denote
frequently the ensemble of 
Minkowski and Grassmann variables by
$X = (x,\theta,\bar{\theta})$.

If we suppose that the fields
$b_{j}, f_{A}$
are free fields and the Hilbert space
${\cal H}$
is in fact the associated Fock space, then one can define in
${\cal H}$
Wick monomials; by multiplication with Grassmann variables we obtain
super-Wick monomials in the extended Fock space
${\cal H}_{G}$.
{\it Super-Wick monomials} are expressions of the type
\bea
:\prod B_{j_{p}}(X_{p})~\prod F_{A_{q}a_{q}}(X_{q})~
\prod \bar{F}_{A_{r}\bar{a}_{r}}(X_{r}):
\label{super-W}
\eea
where some (or all) points can coincide. (Let us note that from these
expressions it follows that super- Wick monomials are causally commuting
as ordinary Wick monomials.) In particular we have a canonical map
$s$
associating to every Wick monomial
$w(x)$
acting in the Hilbert space
${\cal H}$
a super-Wick monomial
$(sw)(x,\theta,\bar{\theta})$
acting in
${\cal H}_{G}$
according to the formula:
\bea
(sw)(x,\theta,\bar{\theta}) \equiv
W_{\theta,\bar{\theta}}~w(x)~W_{\theta,\bar{\theta}}^{-1}
;
\label{s}
\eea
(here $s$ stands for ``sandwich formula" or for ``supersymmetric extension".)

Now we have two elementary results ($[ , ]$ is the graded commutator).
\begin{lemma}
Let us define the operators
\bea
{\cal D}_{a} \equiv {\partial \over \partial \theta^{a}}
- i \sigma^{\mu}_{a\bar{b}} \bar{\theta}^{\bar{b}} \partial_{\mu}
\qquad
\bar{\cal D}_{\bar{a}} \equiv
- {\partial \over \partial \bar{\theta}^{\bar{a}}}
+ i \sigma^{\mu}_{b\bar{a}} \theta^{b} \partial_{\mu}
\label{calD}
\eea
acting on any superfield (or super--Wick polynomials). Then:

(i) the following formul\ae~ are valid for any operator
$T(x,\theta,\bar{\theta})$
acting in
${\cal H}_{G}$:
\bea
{\cal D}_{a} \left[ e^{-\theta \sigma^{\mu}\bar{\theta} P_{\mu}}~
T(x,\theta,\bar{\theta})~
e^{\theta \sigma^{\mu}\bar{\theta} P_{\mu}} \right] =
e^{-\theta \sigma^{\mu}\bar{\theta} P_{\mu}}
\left[ {\partial \over \partial \theta^{a}} T(x,\theta,\bar{\theta}) \right]
e^{\theta \sigma^{\mu}\bar{\theta} P_{\mu}}
\nonumber \\
\bar{\cal D}_{\bar{a}} \left[ e^{-\theta \sigma^{\mu}\bar{\theta} P_{\mu}}
T(x,\theta,\bar{\theta})~
e^{\theta \sigma^{\mu}\bar{\theta} P_{\mu}} \right] =
- e^{-\theta \sigma^{\mu}\bar{\theta} P_{\mu}}
\left[ {\partial \over \partial \bar{\theta}^{\bar{a}}}
T(x,\theta,\bar{\theta}) \right]
e^{\theta \sigma^{\mu}\bar{\theta} P_{\mu}}
\eea
where we use the standard notation
$\zeta \sigma^{\mu} \bar{\theta} \equiv
\zeta^{a} \sigma^{\mu}_{a\bar{b}} \bar{\theta}^{\bar{b}}$.

(ii) for any Wick monomial
$w(x)$
the following relations are true:
\be
{\cal D}_{a} sw = i~s ([Q_{a}, w]), \qquad
\bar{\cal D}_{\bar{a}} sw = i~s([\bar{Q}_{\bar{a}}, w]).
\label{d+s}
\ee
\end{lemma}
\begin{lemma}

(i) Let
$T(x,\theta,\bar{\theta}) = (sw)(x,\theta,\bar{\theta})$.
Then the following formul\ae~ are true:
\bea
W_{\zeta,\bar{\zeta}}~T(x,\theta,\bar{\theta})~W_{\zeta,\bar{\zeta}}^{-1}
= e^{-(\zeta \sigma^{\mu}\bar{\theta} - \theta \sigma^{\mu}\bar{\zeta})
P_{\mu}}
T(x, \theta + \zeta, \bar{\theta} + \bar{\zeta})~
e^{(\zeta \sigma^{\mu}\bar{\theta} - \theta \sigma^{\mu}\bar{\zeta}) P_{\mu}}
\label{BCH}
\eea
\bea
i [ Q_{a}, T(x,\theta,\bar{\theta}) ] = D_{a} T(x,\theta,\bar{\theta}),
\nonumber \\
i [ \bar{Q}_{\bar{a}}, T(x,\theta,\bar{\theta}) ] =
\bar{D}_{\bar{a}} T(x,\theta,\bar{\theta})
\label{BCH-inf}
\eea
where we have defined:
\bea
D_{a} \equiv {\partial \over \partial \theta^{a}}
+ i \sigma^{\mu}_{a\bar{b}} \bar{\theta}^{\bar{b}} \partial_{\mu}
\qquad
\bar{D}_{\bar{a}} \equiv - {\partial \over \partial \bar{\theta}^{\bar{a}}}
- i \sigma^{\mu}_{b\bar{a}} \theta^{b} \partial_{\mu}.
\label{D}
\eea

The formul\ae~(\ref{BCH}) and (\ref{BCH-inf}) are equivalent.

(ii) The operators
$D_{a}$
and
$\bar{D}_{\bar{a}}$
verify the following formul\ae:
\bea
( D_{a} T)^{\dagger} = \pm \bar{D}_{\bar{a}} T^{\dagger},
\nonumber \\
\{D_{a}, D_{b} \} = 0, \quad
\{\bar{D}_{\bar{a}}, \bar{D}_{\bar{b}} \} = 0, \quad
\{D_{a}, \bar{D}_{\bar{b}} \} = -2 i \sigma^{\mu}_{a\bar{b}}~\partial_{\mu}
\label{DD}
\eea
where in the first formula the sign $+ (-)$ correponds to a super-Bose (-Fermi)
field.
\end{lemma}

{\bf Proof:}
The first formula (\ref{BCH}) is a easy consequence of the
Baker-Campbell-Hausdorff formula. One can consider in it that the parameters
$\zeta, \bar{\zeta}$
are ``infinitesimal" and obtain the second formula. The converse statement
follows by recurrence.
$\qed$

For another point of view concerning supersymmetric Hilbert spaces we refer
to the recent paper \cite{Ru}.

\subsection{ Perturbative Supersymmetric Quantum Field Theory\label{axioms}}

We provide here an elementary derivation of the supersymmetric Ward
identities \cite{FL} using the Epstein-Glaser approach to perturbative
quantum fields theory. In this framework one constructs inductively the
chronological products which should satisfy Bogoliubov axioms. We first recall
these axioms for ordinary field theories.

By a
{\it perturbation theory} in the sense of Bogoliubov we mean an ensemble
of
 operator-valued distributions
$
T(w_{1}(x_{1}),\dots,w_{n}(x_{n})) \quad n = 1,2,\dots
$
acting in some Fock space and called {\it chronological products} (where
$
w_{1}(x_{1}),\dots,w_{n}(x_{n})
$
are arbitrary Wick monomials) verifying the following set of axioms:
\begin{itemize}
\item
Skew-symmetry in all arguments
$
w_{1}(x_{1}),\dots,w_{n}(x_{n}):
$
\be
T(\dots,w_{i}(x_{i}),w_{i+1}(x_{i+1}),\dots,) =
(-1)^{f_{i} f_{i+1}} T(\dots,w_{i+1}(x_{i+1}),w_{i}(x_{i}),\dots)
\label{perm}
\ee
where
$f_{i}$
is the number of Fermi fields appearing in the Wick monomial
$w_{i}$.

\item
Poincar\'e invariance: for all
$(a,A) \in inSL(2,\C)$
we have:
\be
U_{a, A} T(w_{1}(x_{1}),\dots,w_{n}(x_{n})) U^{-1}_{a, A} =
T(A\cdot w_{1}(\delta(A)\cdot x_{1}+a),\dots,
A\cdot w_{n}(\delta(A)\cdot x_{n}+a))
\label{invariance}
\ee
where
$A\cdot w
$
is defined through the case
$n = 1$.

Sometimes it is possible to supplement this axiom by corresponding invariance
properties with respect to inversions (spatial and temporal) and charge
conjugation. Also some other global symmetry with respect to some internal
symmetry group
 might be imposed.
\item
Causality: if
$x_{i} \geq x_{j}, \quad \forall i \leq k, \quad j \geq k+1$
then we have:
\be
T(w_{1}(x_{1}),\dots,w_{n}(x_{n})) =
T(w_{1}(x_{1}),\dots,w_{k}(x_{k})) T(w_{k+1}(x_{k+1}),\dots,w_{n}(x_{n})).
\label{causality}
\ee
\item
Unitarity: We define the {\it anti-chronological products} according to
\be
(-1)^{n} \bar{T}(w_{1}(x_{1}),\dots,w_{n}(x_{n})) \equiv \sum_{r=1}^{n}
(-1)^{r} \sum_{I_{1},\dots,I_{r} \in Part(\{1,\dots,n\})}
\epsilon~ T_{I_{1}}(X_{1})\cdots T_{I_{r}}(X_{r})
\label{antichrono}
\ee
where the we have used the notation:
\be
T_{\{i_{1},\dots,i_{k}\}}(x_{i_{1}},\dots,x_{i_{k}}) \equiv
T(w_{i_{1}}(x_{i_{1}}),\dots,w_{i_{k}}(x_{i_{k}}))
\ee
and the sign
$\epsilon$
counts the permutations of the Fermi factors. Then the unitarity axiom is:
\be
\bar{T}(w_{1}(x_{1}),\dots,w_{n}(x_{n}))
= T(w^{*}_{1}(x_{1}),\dots,w^{*}_{n}(x_{n}))
.
\label{unitarity}
\ee
\item
The ``initial condition"
\be
T(w(x)) = w(x).
\label{initial}
\ee
\end{itemize}
\begin{rem}
From (\ref{causality}) one can derive easily that if we have
$x_{i} \sim x_{j}, \quad \forall i \leq k, \quad j \geq k+1$
then:
\be
[ T(w_{1}(x_{1}),\dots,w_{k}(x_{k})), T(w_{k+1}(x_{k+1}),\dots,w_{n}(x_{n}))
]_{\mp} = 0.
\label{commute}
\ee

This relation is essential for the implementation of Epstein-Glaser inductive
construction.
\end{rem}

One extends the definition of
$T(w_{1}(x_{1}),\dots,w_{n}(x_{n}))$
for
$w_{1}(x_{1}),\dots,w_{n}(x_{n})$
Wick polynomials by linearity.

It can be proved that this system of axioms can be supplemented with the
normalization condition of the type
\bea
T(w_{1}(x_{1}),\dots,w_{n}(x_{n}))
\nonumber \\
= \sum \epsilon \quad
<\Omega, T(w_{1}^{'}(x_{1}),\dots,w_{n}^{'}(x_{n}))\Omega>
:w_{1}^{''}(x_{1}),\dots,w_{n}^{''}(x_{n}))
:
\label{wick-chrono2}
\eea
where
$w_{i}^{'}$
and
$w_{i}^{''}$
are Wick submonomials of
$w_{i}$
such that
$w_{i} = :w_{i}^{'} w_{i}^{''}:$
the sign
$\epsilon$
takes care of the permutation of the Fermi fields and
$\Omega$
is the vacuum state; the relation (\ref{wick-chrono2}) is usually called
the Wick expansion property.

We can also include in the induction hypothesis a limitation on the order of
singularity of the vacuum averages of the chronological products associated to
arbitrary Wick monomials
$W_{1},\dots,W_{n}$;
explicitly we have the {\it power counting} formula:
\be
\omega(<\Omega, T(w_{1}(x_{1}),\dots,w_{n}(x_{n}))\Omega>) \leq
\sum_{l=1}^{n} \omega(w_{l}) - 4(n-1)
\label{power}
\ee
where by
$\omega(d)$
we mean the order of singularity of the (numerical) distribution $d$ and by
$\omega(w)$
we mean the canonical dimension of the Wick monomial $w$
We remark here
 that this requirement has important consequences. For instance,
one cannot
 quantize a vector field
$V_{\mu}$
of mass
$m > 0$
imposing the transversality condition \cite{U}, \cite{LW}
\be
\partial_{\mu}~ V^{\mu} = 0
\label{transv}
\ee
because in this case the causal commutator function will have the order
 of
singularity equal to $0$ (instead of $-2$ as for the scalar field). This
behaviour spoils completely the renormalization properties encoded in the
power counting formula. The way out is well-known: one quantize the
vector field without imposing (\ref{transv}) and using an indefinite metric
formalism.

All these axioms have a natural generalization to the case of a supersymmetric
theory. The changes are the following:

\begin{enumerate}
\item
We will make the substitutions
$w_{i} \rightarrow W_{i} = sw_{i}$;
because the expressions
$W_{i}(x,\theta,\bar{\theta})$
depend on the 
Grassmann variables this means that when computing the $S$-matrix
one has 
to integrate over the Grassmann variables too (using of course Berezin
integration).
\item
The computation of the numbers
$f_{i}$
appearing in the symmetry axiom (\ref{perm}) should be made taking into account the
parity of the Grassmann variables also.
\item
The order of singularity should be replaced with the
{\it super-oder of singularity} as defined in \cite{CS} and the canonical
dimension of the superfields should be computed according to additivity and
\bea
\omega(B) = 1, \quad \omega(F) = 3/2, \quad \omega(\partial) = 1.
\nonumber
\eea

These formul\ae~guarantee that the super-order of singularity are identical
with the usual expressions: if the graded commutator of two arbitrary
super-fields is
\be
[ S_{1}(x_{1},\theta_{1},\bar{\theta}_{1}),
S_{2}(x_{2},\theta_{2},\bar{\theta}_{2}) ]
= D_{S_{1},S_{2}}(x_{1},\theta_{1},\bar{\theta}_{1};
x_{2},\theta_{2},\bar{\theta}_{2})
\label{super-CCR}
\ee
then we have
\bea
\omega(D_{S_{1},S_{2}}) \leq \left\{\begin{array}{rcl}
-2 & \mbox{for} & S_{i} \mbox{ super-Bose} \\
-1 & \mbox{for} & S_{i} \mbox{ super-Fermi} \end{array}\right.
\label{super-order}
\eea

In fact, we will see that this formula for the super-order singularity can be
improved for some special choices of the super-fields
$S_{i}$.
\end{enumerate}

All these changes are consistent with the philosophy of supersymmetric
field theory based on  the consistent replacement of the Minkowski space
with the super-space and Wick monomials by super-Wick monomials. It is
the last  assumption which has far-reaching consequences. Indeed this
means that the  Wick expansion property is preserved if and only if the
finite renormalizations are given by quasi-local operators depending
{\bf only on the super-fields} and not on the individual component
fields of the multiplet. The usual proof of the existence of solutions
goes with minimal changes in this supersymmetric setting. We mention
that in this way one can obtain a classification of the theories
according to their renormalizability type as in the usual framework. One
can obtain for instance that the Wess-Zumino model is
super-normalizable in this sense \cite{CS}.

Let us define the operators
$D_{a}^{l}, \quad \bar{D}_{\bar{a}}^{l}, \quad l = 1,\dots,n$
by the formul\ae~(\ref{D}) associated to the corresponding variable
$X_{l}, \quad l= 1,\dots,n$.
Then we have the following result:
\begin{thm}
Suppose that the expression
$T(X)$
verifies the identities (\ref{BCH-inf}). Then one can choose the chronological
products
$T(X_{1},\dots,X_{n}) \equiv T(W_{1}(X_{1}),\dots,W_{n}(X_{n}))$
such that, beside the preceding axioms, the following identities are verified:
\bea
i [ Q_{a}, T(X_{1},\dots,X_{n}) ]
= \sum_{l=1}^{n} D^{l}_{a} T(X_{1},\dots,X_{n}),
\nonumber \\
i [ \bar{Q}_{\bar{a}}, T(X_{1},\dots,X_{n}) ] =
\sum_{l=1}^{n} \bar{D}^{l}_{\bar{a}} T(X_{1},\dots,X_{n});
\label{susy-ward}
\eea

(here
$[ , ]$
is the graded commutator).
\end{thm}

{\bf Proof:}
Goes by induction. For
$n = 1$
the identities are valid by hypothesis. Suppose that we have
\bea
i [ Q_{a}, T(X_{1},\dots,X_{p}) ]
= \sum_{l=1}^{p} D^{l}_{a} T(X_{1},\dots,X_{p}),
\nonumber \\
i [ \bar{Q}_{\bar{a}}, T(X_{1},\dots,X_{p}) ] =
\sum_{l=1}^{p} \bar{D}^{l}_{\bar{a}} T(X_{1},\dots,X_{p})
\eea
for
$p = 1,\dots,n-1$.
Then one can easily prove using causality that in order $n$ we have:
\bea
i [ Q_{a}, T(X_{1},\dots,X_{n}) ]
= \sum_{l=1}^{n} D^{l}_{a} T(X_{1},\dots,X_{n}) + P_{a}(X_{1},\dots,X_{n})
\nonumber \\
i [ \bar{Q}_{\bar{a}}, T(X_{1},\dots,X_{n}) ] =
\sum_{l=1}^{n} \bar{D}^{l}_{\bar{a}} T(X_{1},\dots,X_{n})
+ \bar{P}_{\bar{a}}(X_{1},\dots,X_{n})
\eea
where
$P_{a}, \quad \bar{P}_{\bar{a}}$
are quasi-local operators: the supersymmetric anomalies. There are a number
of restrictions on these anomalies. First we have
\be
P_{a}(X_{1},\dots,X_{n})^{\dagger} = \bar{P}_{\bar{a}}(X_{1},\dots,X_{n})
\ee
which follows from the unitarity axiom. Next we have Wess-Zumino consistency
relations which follow by considering the (graded) Jacobi identities:
\bea
[ Q_{a}, [ Q_{b} , T(X_{1},\dots,X_{n}) ] ] = - (a \leftrightarrow b),
\nonumber \\
~[ Q_{a}, [ \bar{Q}_{\bar{b}} , T(X_{1},\dots,X_{n}) ] ]
- [ \bar{Q}_{\bar{b}}, [T(X_{1},\dots,X_{n}), Q_{a} ] ]
= 2~i~\sigma^{\mu}_{a\bar{b}} \sum_{l=1}^{n}
\partial^{l}_{\mu} T(X_{1},\dots,X_{n});
\eea
(here
$[ \cdot,\cdot ]$
is the graded commutator). If we substitute here the preceding relations we
immediately get:
\bea
[ Q_{a}, P_{b}(X_{1},\dots,X_{n}) ]
+ i \sum_{l=1}^{n} D^{l}_{b} P_{a}(X_{1},\dots,X_{n})
= - (a \leftrightarrow b)
\nonumber \\
~[ Q_{a}, \bar{P}_{\bar{b}}(X_{1},\dots,X_{n}) ]
+ [ \bar{Q}_{\bar{b}}, P_{a}(X_{1},\dots,X_{n}) ]
\nonumber \\
+ i \sum_{l=1}^{n} [ D^{l}_{a} \bar{P}_{\bar{b}}(X_{1},\dots,X_{n})
+ \bar{D}^{l}_{\bar{b}} P_{a}(X_{1},\dots,X_{n}) ] = 0
\label{WZ}
\eea

It is now a straightforward but rather long computation to obtain a generic
form of the supersymmetric anomalies and to show that they  can be eliminated
by conveniently redefining the chronological products. Let us give the details.
It is convenient to work in new Grassmann variables:
\bea
\Theta \equiv {1\over n}~\sum_{l=1}^{n} \theta_{l}, \qquad
\zeta_{l} \equiv \theta_{l} - \theta_{n}, \quad l = 1,\dots,n - 1.
\nonumber
\eea
In these new variables we have:
\bea
D_{a} = {\partial \over \partial \Theta^{a}} + d_{a}, \qquad
\bar{D}_{\bar{a}} = {\partial \over \partial \bar{\Theta}^{\bar{a}}}
+ \bar{d}_{\bar{a}}
\eea
where the operators
$d_{a}$
and
$\bar{d}_{\bar{a}}$
involve only the variables
$\zeta, \bar{\zeta}$.

The generic form of the anomaly is:
\bea
P_{a} = p_{a} + \Theta^{c} p_{ac} + \bar{\Theta}^{\bar{c}} p_{a\bar{c}}
+ (\bar{\Theta}\bar{\Theta}) p'_{a} + (\Theta \Theta) p^{"}_{a}
+ \Theta^{c} \bar{\Theta}^{\bar{d}} p_{ac\bar{d}}
\nonumber \\
+ (\bar{\Theta}\bar{\Theta}) \Theta^{c} p'_{ac}
+ (\Theta\Theta) \bar{\Theta}^{\bar{c}} p'_{a\bar{c}}
+ (\bar{\Theta}\bar{\Theta}) (\Theta\Theta) p^{"}_{a}
\label{anomaly}
\eea
where the expressions
$p^{\dots}_{a\dots}$
do not depend on
$\Theta, \quad \bar{\Theta}$;
we use the usual notations
$
\bar{\theta}\bar{\theta} \equiv \bar{\theta}_{\bar{a}} \bar{\theta}^{\bar{a}}
$
and
$
\theta\theta \equiv \theta^{a}\theta_{a}
$
for any Grassmann variable
$\theta$.
We want to prove that this anomaly is in fact a coboundary i.e. of the form:
\bea
(\delta P)_{a} = i [ Q_{a}, P ] - \sum_{l=1}^{n} D_{a}^{l} P
= i [ Q_{a}, P ]
- \left( {\partial \over \partial \Theta^{a}} + d_{a}
 \right) P
\label{coboundary}
\eea
where $\delta$ is the cochain operator and $P$ is arbitrary. Now we make a
succession of finite renormalizations of the type
\bea
T(X_{1},\dots,X_{n}) \quad \rightarrow \quad T(X_{1},\dots,X_{n})
+ P_{i}(X_{1},\dots,X_{n}), \quad i = 1,\dots,6
\label{finite-renorm}
\eea
and we will find
$P_{a} = 0$
as a result.
\begin{itemize}
\item
We define
\bea
P_{1} \equiv \Theta^{a} p_{a}
\nonumber
\eea
and (\ref{finite-renorm}) makes
$p_{a} = 0$
in (\ref{anomaly}) and
$p_{ab}$
is redefined.
\item
Then we get from the first equation (\ref{WZ}) that
$p_{ab} = - (a \leftrightarrow b)$
i.e.
$p_{ab} = \epsilon_{ab} p$.
We define
\bea
P_{2} = {1\over 2} (\Theta\Theta) p
\nonumber
\eea
and (\ref{finite-renorm}) makes
$p = 0$;
the first equation (\ref{WZ}) gives now
$p^{"}_{a} = 0$
in (\ref{anomaly}).
\item
We define
\bea
P_{3} =  \Theta^{b} \bar{\Theta}^{\bar{c}} p_{b\bar{c}}
\nonumber
\eea
and (\ref{finite-renorm}) makes
$p_{b\bar{c}} = 0$
in (\ref{anomaly}).
\item
We define
\bea
P_{4} = (\bar{\Theta}\bar{\Theta}) \Theta^{c} p'_{c}
\nonumber
\eea
and (\ref{finite-renorm}) makes
$p'_{a} = 0$
in (\ref{anomaly});
$p'_{ac} = 0$
is redefined.
\item
The first equation (\ref{WZ}) gives
$p_{ab\bar{c}} = - (a \leftrightarrow b)$
i.e.
$p_{ab\bar{c}} = \epsilon_{ab} \bar{p}_{\bar{c}}$.
We define
\bea
P_{5} = {1\over 2} (\Theta\Theta) \bar{\Theta}^{\bar{c}} \bar{p}_{\bar{c}}
\nonumber
\eea
and (\ref{finite-renorm}) makes
$p_{ab\bar{c}} = 0$;
The first equation (\ref{WZ}) gives now
$p'_{a\bar{d}} = 0$.
\item
The first equation (\ref{WZ}) gives
$p'_{ab} = - (a \leftrightarrow b)$
i.e.
$p'_{ab} = \epsilon_{ab} p'$.
We define
\bea
P_{6} = {1\over 2} (\bar{\Theta}\bar{\Theta}) (\Theta\Theta) p'
\nonumber
\eea
and (\ref{finite-renorm}) makes
$p'_{ab} = 0$.
\end{itemize}

In conclusion, the first equation (\ref{WZ}) can be used to fix the form
of the anomaly to:
\bea
P_{a} = (\bar{\Theta}\bar{\Theta}) (\Theta\Theta) p^{"}_{a}.
\nonumber
\eea

Now the second equation (\ref{WZ}) gives immediately
$p^{"}_{a} = 0$
and we finally obtain
$P_{a} = 0$.
$\qed$

We will call the identities (\ref{susy-ward}) from the statement of the
theorem the 
{\it supersymmetric normalization conditions}. Now we have the
following
\begin{cor}
In the conditions of the preceding theorem we have
\bea
\sum_{l=1}^{n} D^{l}_{a} <\Omega,T(X_{1},\dots,X_{n}) \Omega> = 0,
\nonumber \\
\sum_{l=1}^{n} \bar{D}^{l}_{\bar{a}} <\Omega, T(X_{1},\dots,X_{n}) \Omega> = 0.
\label{susy-ward-vacuum}
\eea

If we go into the momentum space, the supersymmetric normalization conditions
derived
 above become:
\bea
\sum_{l=1}^{n} \left( {\partial \over \partial \theta^{a}_{l}}
- \sigma^{\mu}_{a\bar{b}} \bar{\theta}^{\bar{b}}_{l} p_{\mu}^{l} \right)
\tilde{t}(p_{1},\theta_{1},\bar{\theta}_{1};\dots;
p_{n},\theta_{n},\bar{\theta}_{n}) = 0,
\nonumber \\
\sum_{l=1}^{n} \left( {\partial \over \partial \bar{\theta}^{\bar{a}}}_{l}
- \sigma^{\mu}_{b\bar{a}} \theta^{b}_{l} p_{\mu}^{l} \right)
\tilde{t}(p_{1},\theta_{1},\bar{\theta}_{1};\dots;
p_{n},\theta_{n},\bar{\theta}_{n}) = 0
\label{momentum}
\eea
where
\be
\tilde{t}(p_{1},\theta_{1},\bar{\theta}_{1};\dots;
p_{n},\theta_{n},\bar{\theta}_{n}) \equiv
{1 \over (2\pi)^{2n}} \int dx_{1} \dots dx_{n}
\exp(-i \sum_{l=1}^{n} p_{l} \cdot x_{l})
<\Omega, T(X_{1},\dots,X_{n}) \Omega>.
\ee

Moreover, these identities are equivalent to (\ref{susy-ward}).
\end{cor}

{\bf Proof:}
The implication (\ref{susy-ward}) $\Longrightarrow$ (\ref{susy-ward-vacuum})
is elementary: one simply takes the vacuum average and uses the first relation
of (\ref{vac}). The converse implications follows if one uses the Wick
expansion property (\ref{wick-chrono2}) and (\ref{BCH-inf}).
$\qed$

Let us note that the equations (\ref{momentum}) can be ``integrated" to
the usual form of the supersymmetric ``Ward identities" \cite{FL}:
\begin{cor}
The following identity is valid for any Grassmann variables
$\zeta,~\bar{\zeta}$:
\bea
\tilde{t}(p_{1},\theta_{1} + \zeta,\bar{\theta}_{1}+ \bar{\zeta};\dots;
p_{n},\theta_{n}+ \zeta,\bar{\theta}_{n}+ \bar{\zeta})
\nonumber \\
= \exp\left[ \sum_{l=1}^{n} \left( \zeta \sigma^{\mu} \bar{\theta}_{l}
- \theta_{l} \sigma^{\mu} \bar{\zeta} \right) p_{\mu}^{l} \right] \quad
\tilde{t}(p_{1},\theta_{1},\bar{\theta}_{1};\dots;
p_{n},\theta_{n},\bar{\theta}_{n})
\label{ward-FL}
\eea
\end{cor}

{\bf Proof:}
We write the two identities (\ref{momentum}) from the preceding Corollary in
a compact form: we denote
\be
c \equiv \sum_{l=1}^{n} \left( \zeta \sigma^{\mu} \bar{\theta}_{l}
-  \theta_{l} \sigma^{\mu} \bar{\zeta} \right) p_{\mu}^{l}
\qquad
D \equiv \sum_{l=1}^{n}
\left( \zeta^{a} {\partial \over \partial \theta^{a}_{l}}
+ \bar{\zeta}^{\bar{a}} {\partial \over \partial \bar{\theta}^{\bar{a}}_{l}}
\right);
\ee
here $c$ is a Grassmann number and $D$ is a differential operator. Then the
identities (\ref{momentum}) can be written in the very compact way:
\be
D \tilde{t}(p_{1},\theta_{1},\bar{\theta}_{1};\dots;
p_{n},\theta_{n},\bar{\theta}_{n})
= c \tilde{t}(p_{1}\theta_{1},\bar{\theta}_{1};\dots;
p_{n},\theta_{n},\bar{\theta}_{n}).
\ee

If we note the identity
\be
D~ c = 0
\ee
we immediately iterate the preceding identity to:
\be
\sum_{k=0}^{4} {1\over k!} D^{k}
\tilde{t}(p_{1},\theta_{1},\bar{\theta}_{1};\dots;
p_{n},\theta_{n},\bar{\theta}_{n})
= e^{c} \tilde{t}(p_{1}\theta_{1},\bar{\theta}_{1};\dots;
p_{n},\theta_{n},\bar{\theta}_{n}).
\ee
Now we use Taylor formula for superfunctions:
\be
\sum_{k=0}^{4} {1\over k!} D^{k} f(\theta, \bar{\theta}) =
f(\theta + \zeta, \bar{\theta} + \bar{\zeta})
\ee
and obtain the formula from the statement.
$\qed$

This corollary leads to:
\begin{cor}
The most general form of
$\tilde{t}$
is:
\bea
\tilde{t}(p_{1},\theta_{1},\bar{\theta}_{1};\dots;
p_{n},\theta_{n},\bar{\theta}_{n})
\nonumber \\
= \exp\left[ \sum_{l=1}^{n-1} \left( \theta_{n} \sigma^{\mu} \bar{\theta}_{l}
- \theta_{l} \sigma^{\mu} \bar{\theta_{n}} \right) p_{\mu}^{l} \right] \quad
\tilde{t}_{0}(p_{1},\theta_{1n},\bar{\theta}_{1n};\dots;
p_{n-1},\theta_{n-1,n},\bar{\theta}_{n-1,n})
\label{general}
\eea
where
$t_{0}$
is an arbitrary distribution; here
$\theta_{ij} \equiv \theta_{i} - \theta_{j}$.
\end{cor}
{\bf Proof:}
We simply take
$\zeta = - \theta_{n}, \quad \bar{\zeta} = - \bar{\theta}_{n}$
in the formula from the preceding corollary and take into account that
conservation of the momentum restricts the support of
$\tilde{t}$
to the subset
$\sum_{l=1}^{n} p_{l} = 0$.
We obtain the formula from the statement for a certain
$t_{0}$
and then we show that the formula (\ref{general}) identically verifies
(\ref{ward-FL}).
$\qed$

The importance of the formul\ae~(\ref{susy-ward-vacuum}) follows from the
fact that it drastically limits the possible finite renormalizations.
Indeed, we have:
\begin{prop}
(i) Suppose that the chronological products verify the supersymmetric Ward
identities (\ref{susy-ward-vacuum})
. Then the most general arbitrariness of
these products are of the
 form
\be
\sum d_{i}(X_{1},\dots,X_{n})  W_{i}(X_{1},\dots,X_{n})
\ee
where
$W_{i}$
are super-Wick monomials (\ref{super-W}) and
$d_{i}$
are distributions with the support in the diagonal set
$x_{1} = \cdots = x_{n}$,
depending on the Grassmann variables in such a way that one has
\be
\sum_{l=1}^{n} D_{a} d_{i} = 0, \qquad
\sum_{l=1}^{n} \bar{D}_{\bar{a}} d_{i} = 0 \qquad \forall i.
\ee

(ii) The general form of such a distribution $d$ is in $p$-space:
\bea
d(p_{1},\theta_{1},\bar{\theta}_{1};\dots;
p_{n},\theta_{n},\bar{\theta}_{n})
\nonumber \\
= \exp\left[ \sum_{l=1}^{n-1} \left( \theta_{n} \sigma^{\mu} \bar{\theta}_{l}
- \theta_{l} \sigma^{\mu} \bar{\theta_{n}} \right) p_{\mu}^{l} \right] \quad
d_{0}(p_{1},\theta_{1n},\bar{\theta}_{1n};\dots;
p_{n-1},\theta_{n-1,n},\bar{\theta}_{n-1,n})
\eea
where
$d_{0}$
is a polynomial in the momenta
$p_{i}, \quad i = 1,\dots,n-1$.
\end{prop}

{\bf Proof:}
(i) It follows from the Wick expansion property for superfields and the
supersymmetric normalization conditions (\ref{susy-ward-vacuum}).

(ii) Follows from the formula (\ref{general}).
$\qed$

It is natural in this context to define the {\it supersymmetric} $\delta$
{\it distribution} to correspond to
$t_{0} = 1$:
\bea
\tilde{\delta}_{S}(p_{1},\theta_{1},\bar{\theta}_{1};\dots;
p_{n},\theta_{n},\bar{\theta}_{n})
= \exp\left[ \sum_{l=1}^{n-1} \left( \theta_{n} \sigma^{\mu} \bar{\theta}_{l}
- \theta_{l} \sigma^{\mu} \bar{\theta_{n}} \right) p_{\mu}^{l} \right]
\eea
(in momentum space) and
\bea
\delta_{S}(X_{1},\dots,X_{n})
= \exp\left[ - i~\sum_{l=1}^{n-1}
\left( \theta_{n} \sigma^{\mu} \bar{\theta}_{l}
- \theta_{l} \sigma^{\mu} \bar{\theta_{n}} \right) \partial_{\mu}^{l} \right]
\delta(x_{1} - x_{n}) \dots \delta(x_{n-1} - x_{n})
\eea
(in coordinate space). Let us note that we have:
\be
\delta_{S}(X_{1};\dots;X_{n}) = \prod_{l=1}^{n-1} \delta_{S}(X_{l};X_{n})
\ee
which has a well-known analogue for ordinary distributions.

We can re-express the formul\ae~(\ref{momentum}) and (\ref{ward-FL}) in
coordinate space if we define \cite{O} for any test function
$f \in {\cal S}(\R^{4})$
and any
$\theta^{\mu} \in {\cal G}$
\be
f(x+ \theta) \equiv
\sum_{k=0}^{2} {1\over k!} \theta^{\alpha} \partial_{\alpha}^{k} f(x)
\equiv \sum_{k=0}^{2} \sum_{\mu_{1} \leq \cdots \leq \mu_{k}}
\theta^{\mu_{1}} \dots \theta^{\mu_{k}}
\partial_{\mu_{1}} \dots \partial_{\mu_{n}} f(x)
\ee
and similarly for
$\bar{\theta}$.
Then:
\begin{cor}
One can choose the chronological products such that we have:
\bea
<\Omega,T(x_{1},\theta_{1},\bar{\theta}_{1};\dots;
x_{n},\theta_{n},\bar{\theta}_{n})\Omega>
\nonumber \\
= <\Omega,T(x_{1} + c(\zeta,\theta_{1}),\theta_{1} + \zeta,
\bar{\theta}_{1} + \bar{\zeta};\dots;
x_{n} + c(\zeta,\theta_{n}),\theta_{n} + \zeta,\bar{\theta}_{n} + \bar{\zeta})
\Omega>
\eea
where
$
c^{\mu}(\zeta,\theta) \equiv
- i~(\zeta \sigma^{\mu} \bar{\theta} - \theta \sigma^{\mu} \bar{\zeta})
$
or, in the integrated form:
\bea
W_{\theta,\bar{\theta}}
T(x_{1},\theta_{1},\bar{\theta}_{1};\dots;x_{n},\theta_{n},\bar{\theta}_{n})
W_{\theta,\bar{\theta}}^{-1}
\nonumber \\
= T(x_{1} + c(\zeta,\theta_{1}),\theta_{1} + \zeta,
\bar{\theta}_{1} + \bar{\zeta};\dots;
x_{n} + c(\zeta,\theta_{n}),\theta_{n} + \zeta,\bar{\theta}_{n} + \bar{\zeta}).
\eea
\end{cor}

For the analysis of supersymmetric anomalies in the traditional approach
to SUSY, based on BRST quantization, we refer to \cite{Br}, \cite{Di},
\cite{HLW}.

We close by mentioning that the usual improved formul\ae~ for the order
of singularity (see for instance \cite{Ca}, \cite{CL}, \cite{De},
\cite{FL},
 for the Wess-Zumino model) can be rigorously justified if
one
 uses another normalization condition. The starting point are the
formul\ae~ (\ref{d+s}
); if
$W(x,\theta,\bar{\theta})$
is a supersymmetric Wick monomial, let us define the operation of
restriction to the ``initial value" (in the Grassmann variables):
\be
(rW)(x) \equiv W(x,0,0).
\label{r}
\ee
Then the formul\ae~ (\ref{d+s}) imply
\be
{\cal D}_{a} W = i~s ([Q_{a}, rW]), \qquad
\bar{\cal D}_{\bar{a}} W = i~s([\bar{Q}_{\bar{a}}, rW]).
\label{s+r}
\ee

It is clear that these equations can be regarded as a system of  partial
differential equations (in the Grassmann variables) and this system
determines uniquely the supersymmetric Wick monomials $W$ if one knows
the ``initial values"
$w = rW$.
(Indeed if there are two solutions, then their difference verifies the
associated homogeneous equation which tells that there is no dependence on
the Grassmann variables; but the ``initial values" for the difference is zero.)

One can promote equations of this type as new supersymmetric
normalization conditions. Let us define the operators
${\cal D}_{a}^{l}, \quad \bar{\cal D}_{\bar{a}}^{l}, \quad l = 1,\dots,n$
by the formul\ae~(\ref{calD}) associated to the corresponding variable
$X_{l}, \quad l= 1,\dots,n$.
Then we have a new normalization condition (which seems to be new in the
literature) contained in the following
\begin{thm}
Let
$W_{1}(X_{1}),\dots,W_{n}(X_{n})$
be some supersymmetric Wick monomials. Then one can normalize the chronological
products
$T(W_{1}(X_{1}),\dots,W_{n}(X_{n}))$
such that the following identities are verified
$\forall l =1,\dots,n$:
\bea
{\cal D}^{l}_{a}  T(W_{1}(X_{1}),\dots,W_{n}(X_{n})) = i
T(W_{1}(X_{1}),\dots,s([Q_{a}, rW_{l}])(X_{l}),\dots,W_{n}(X_{n}))
\nonumber \\
\bar{\cal D}^{l}_{\bar{a}}  T(W_{1}(X_{1}),\dots,W_{n}(X_{n})) = i
T(W_{1}(X_{1}),\dots,s([\bar{Q}_{\bar{a}}, rW_{l}])(X_{l}),
\dots,W_{n}(X_{n}))
\label{susy-map}
\eea
\label{super}
\end{thm}

{\bf Proof:}
We denote
$w_{j} \equiv rW_{j}, \quad j = 1,\dots,n$;
these are ordinary Wick monomials and we can choose a solution of the
Bogoliubov axioms
$T(w_{1}(x_{1}),\dots,w_{n}(x_{n}))$
acting in the Hilbert space
${\cal H}$.
Then we {\bf define}
$T(W_{1}(X_{1}),\dots,W_{n}(X_{n}))$
as the (unique) solution of the system of equations from the statement of the
theorem with the ``initial conditions"
$T(w_{1}(x_{1}),\dots,w_{n}(x_{n})).$
(The uniqueness argument is the same as above). It remains to show that this
solution also verifies Bogoliubov axioms. For the ``initial condition"
axiom (\ref{initial}) this is trivial: the system (\ref{susy-map}) goes into
(\ref{s+r}) for
$n = 1$.
The causality axiom follows again from an unicity argument. Indeed, suppose
that we have
$x_{i} \geq x_{j}, \quad \forall i \leq k, \quad \forall j \geq k+1$.
We want to prove that the expression
\bea
T^{\prime}(W_{1}(X_{1}),\dots,W_{n}(X_{n})) \equiv
\nonumber \\
T(W_{1}(X_{1}),\dots,W_{n}(X_{n})) -
T(W_{1}(X_{1}),\dots,W_{k}(X_{k})) T(W_{k+1}(X_{k+1}),\dots,W_{n}(X_{n}))
\nonumber
\eea
is null. For this one notices that the expressions
$T^{\prime}$
also verifies the system (\ref{susy-map}) and it has the ``initial condition"
equal to $0$ because we have by assumption
\bea
T(w_{1}(x_{1}),\dots,w_{n}(x_{n})) =
T(w_{1}(x_{1}),\dots,w_{k}(x_{k})) T(w_{k+1}(x_{k+1}),\dots,w_{n}(x_{n})).
\nonumber
\eea

Then the unicity argument gives us
$T^{\prime} = 0$.
In the same way one checks the validity of unitarity and of Lorentz covariance.
$\qed$

We can use this new normalization condition to limit drastically the
arbitrariness of the chronological products. Indeed, if such a normalization
of the chronological products is adopted then it follows that the arbitrariness
is contained into the arbitrariness of the ``initial value" chronological
products
$T(w_{1}(x_{1}),\dots,w_{n}(x_{n}))$.
For instance, if we consider the supersymmetric interaction
\be
T(x,\theta,\bar{\theta}) = :\Phi(x,\theta,\bar{\theta})^{3}:
+ :(\Phi(x,\theta,\bar{\theta})^{\dagger})^{3}:
\label{phi3}
\ee
it follows that the ``initial value" chronological products correspond to a
$\phi^{3}$-theory
which is known to be super-normalizable. However, the preceding interaction
does not correspond to the Wess-Zumino model! According to \cite{CS} the
interaction for this model is:
\be
T(x,\theta,\bar{\theta}) = \delta(\theta):\Phi(x,\theta,\bar{\theta})^{3}:
+ \delta(\bar{\theta}) :(\Phi(x,\theta,\bar{\theta})^{\dagger})^{3}:
\label{wz-phi3}
\ee
where
$\delta(\theta) \equiv \theta\theta$
and
$\delta(\bar{\theta}) \equiv \bar{\theta}\bar{\theta}$.
Because of the presence of these Grassmann coefficients, we cannot impose
the normalization conditions from the preceding theorems.  However, we
can extend the definition of the chronological products through
linearity
\bea
T(\sum f_{i_{1}}(\theta_{1},\bar{\theta}_{1}) W_{i_{1}}(X_{1}),\dots,
\sum f_{i_{n}}(\theta_{n},\bar{\theta}_{n}) W_{i_{n}}(X_{n}))
\nonumber \\
\equiv \sum \pm f_{i_{1}}(\theta_{1},\bar{\theta}_{1}) \dots,
f_{i_{n}}(\theta_{n},\bar{\theta}_{n})\quad
T(W_{i_{1}}(X_{1}),\dots,W_{i_{n}}(X_{n}))
\nonumber
\eea
(where the signs takes care of permutations of odd Grassmann factors); the
expressions so defined verify also Bogoliubov axioms. However, these new
chronological products will not verify the normalization conditions
appearing in the two theorems proved above. Nevertheless, in this way we
will obtain the chronological products for the Wess-Zumino model as linear
combinations of the chronological products of the super-normalizable
$:\Phi^{3}:$
model (\ref{phi3}).

We close remarking that it is not clear if both normalization condition
(\ref{susy-map}) and (\ref{susy-ward}) can be implemented as the same time.

\section{Explicit Construction of Supersymmetric Multiplets\label{multiplets}}

In this Section we construct the basic multiplets which will be used to
build the supersymmetric extension of gauge theory: a vector multiplet,
a pair of fermionic ghost and anti-ghost multiplets and a bosonic ghost
multiplet. We will start with the last multiplet because it is in fact a
Wess-Zumino multiplet. For completeness we present the derivation from
\cite{LW}. Finally we will connect these multiplets by a gauge charge operator
in strict analogy with the usual construction of quantum gauge theory
\cite{Sc}. This then shows the usefulness of these multiplets.

\subsection{The Wess-Zumino Multiplet\label{wz-multiplet}}

The most simple case of the general framework described in Section
\ref{susy-multiplets} is when all bosonic fields are real scalar
$\phi^{(j)}, \quad j = 1,\dots,s$
and all Fermionic fields are Majorana spin $1/2$ fields
$f^{(A)}_{a}, A = 1,\dots,f$.

As always in $S$-matrix theory we are dealing with free fields: the scalar
(resp. Majorana) fields verify Klein-Gordon (resp. Dirac) equations:
\bea
(\partial^{2} + m_{j}^{2}) \phi^{(j)} = 0, \quad j = 1,\dots,s
\nonumber \\
i~\sigma^{\mu}_{a\bar{b}} \partial_{\mu} \bar{f}^{(A)\bar{b}}
= M_{A} f^{(A)}_{a}, \qquad
- i~\sigma^{\mu}_{a\bar{b}} \partial_{\mu} f^{(A)a}
= M_{A} \bar{f}^{(A)}_{\bar{b}}, \quad A = 1,\dots,f
\label{kg+dirac}
\eea
and convenient causal (anti)commutation relations.

For later convenience we introduce the (diagonal) mass matrices
$m \in M_{\R}(s,s), \quad M \in M_{\R}(f,f)$
according to:
\be
m_{jk} \equiv \delta_{jk} m_{j}, \qquad
M_{AB} \equiv \delta_{AB} M_{A}.
\ee

In this case a classification theorem is available \cite{LW}. First we
remind the reader the definition of {\it Wess-Zumino multiplet} \cite{WZ}. It
corresponds to the case
$f = 1$
and
$s = 2$.
Then we can consider that we have in fact a complex scalar field
$\phi$
and a spin $1/2$ Majorana field
$f_{a}$
of the same mass $m$. The relations (\ref{tensor}) are in this case by
definition:
\bea
[Q_{a}, \phi ] = 0, \qquad [\bar{Q}_{\bar{a}}, \phi^{\dagger} ] = 0
\nonumber \\
i~[Q_{a}, \phi^{\dagger} ] = 2 f_{a}, \qquad
i~[\bar{Q}_{\bar{a}}, \phi] = 2 \bar{f}_{\bar{a}}
\nonumber \\
~\{ Q_{a}, f_{b} \} = - i~m~\epsilon_{ab} \phi, \qquad
\{ \bar{Q}_{\bar{a}}, \bar{f}_{\bar{b}} \} = i~m~\epsilon_{\bar{a}\bar{b}}
\phi^{\dagger}
\nonumber \\
~\{ Q_{a}, \bar{f}_{\bar{b}} \} = \sigma^{\mu}_{a\bar{b}} \partial_{\mu}\phi
,
\qquad
\{ \bar{Q}_{\bar{a}}, f_{b} \} = \sigma^{\mu}_{b\bar{a}}
\partial_{\mu}\phi^{\dagger}
.
\label{wess}
\eea

The first vanishing commutators are also called (anti) chirality condition.
The causal (anti)commutators are:
\bea
\left[ \phi(x), \phi(y)^{\dagger} \right] = - 2i~ D_{m}(x-y),
\nonumber \\
\left\{ f_{a}(x), f_{b}(y) \right\} = i~~\epsilon_{ab}~m~D_{m}(x-y),
\nonumber \\
\left\{ f_{a}(x), \bar{f}_{\bar{b}}(y) \right\}
= \sigma^{\mu}_{a\bar{b}}~\partial_{\mu}D_{m}(x-y)
\label{CCR-wess}
\eea
and the other (anti)commutators are zero.
\begin{rem}
The supersymmetry invariance manifests itself in a very nice
way connecting the causal (anti)commutators from the bosonic and the
fermionic sector; the last two relations in the preceding formula
are completely determined via the consistency relations (\ref{susy+CCR}).
\end{rem}

One can easily prove that all consistency conditions (\ref{lorentz}),
(\ref{susy+lorentz}), (\ref{factor}) and (\ref{susy+CCR}) are verified. Then
we have the following result from \cite{WZ}; we will provide the proof
because the argument proves to be rewarding for more complicated cases.
\begin{thm}
Let
$\phi^{(j)}, \quad j = 1,\dots,s$
be bosonic Hermitian fields
and
$f^{(A)}_{a}, A = 1,\dots,f$.
fermionic Majorana fields of spin $1/2$ fields:
\be
(\phi^{(j)})^{\dagger} = \phi^{(j)}, \quad j = 1,\dots,s \qquad
(f^{(A)}_{a})^{\dagger} = \bar{f}^{(A)}_{\bar{a}},
\quad A = 1,\dots,f, \quad a=1,2
\ee
forming a supersymmetric multiplet. Then we necessarily have
$s = 2f$
and the multiplet is a direct sum of irreducible multiples of the Wess-Zumino
type.
\end{thm}

{\bf Proof:}
According to the hypothesis, the generic form of (\ref{tensor}) must be:
\bea
i~[Q_{a}, \phi^{(j)} ] = \sum_{A=1}^{f} {\cal A}_{1}^{jA} f^{(A)}_{a}
\nonumber \\
\{ Q_{a}, f^{(A)}_{b} \} = \epsilon_{ab} \sum_{j=1}^{s} {\cal A}_{2}^{Aj}
\phi^{(j)}
\qquad
\{ Q_{a}, \bar{f}^{(A)}_{\bar{b}} \} = \sigma^{\mu}_{a\bar{b}}
\sum_{j=1}^{s} {\cal A}_{3}^{Aj} \partial_{\mu}\phi^{(j)}
\label{ansatz}
\eea
where
${\cal A}_{l}, \quad l =1,2,3$
are some complex matrices. Remark that in writing down such an ansatz we have
taken into account the Lorentz covariance restriction. We also remark that
higher derivatives can be eliminated in the right hand side if one uses
the equations of motion. Using the hypothesis that the scalar (resp. spinor)
fields are Hermitian (resp. Majorana) we obtain from the preceding relation
another three similar relations. Now we put to use the consistency conditions;
it is elementary to obtain from (\ref{susy+lorentz}):
\bea
{\cal A}_{2} {\cal A}_{1} = 0,  \qquad {\cal A}_{3} {\cal A}_{1} = 0,
\qquad {\cal A}_{2} \bar{\cal A}_{1} = - 2~i~M
\nonumber \\
\bar{\cal A}_{1} {\cal A}_{3} + {\cal A}_{1} \bar{\cal A}_{3} = 2 I_{s},
\qquad
\bar{\cal A}_{3} {\cal A}_{1} = 2 I_{f}
\label{wz1}
\eea
where
$I_{s}$
(resp.
$I_{f}$)
is the identity matrix in $s$ (resp. $f$) dimensions and the bar denotes
complex conjugation and from the consistency (\ref{factor}) with the equation
of motion:
\bea
{\cal A}_{2} = -~i~M~{\cal A}_{3}, \qquad
M~{\cal A}_{2} = -~i~{\cal A}_{3}~m^{2}, \qquad
m^{2}~{\cal A}_{1} = {\cal A}_{1}~M^{2}.
\label{wz2}
\eea

If we take the trace of the last two relations (\ref{wz1}) we get
$s = 2f$. Next, we define the matrices
$A, B \in M_{\R}(s,s)$
according to the formula:
\be
A^{T} \equiv ({\cal A}_{3}, \bar{\cal A}_{3} ), \qquad
B \equiv (\bar{\cal A}_{1}, {\cal A}_{1} )
\label{AB}
\ee
where the bracket means juxtaposition of rectangular matrices. Then we easily
find out from (\ref{wz1}) that
\be
A~B = B~A = 2~I_{s};
\ee
this means that the matrices $A$ and $B$ are invertible. It follows that we
can replace the $s$ real fields
$\phi^{(j)}$
by
$f = s/2$
complex fields according to:
\be
\phi^{(A)} \equiv \sum_{j=1}^{s} {\cal A}^{Aj}_{3} \phi^{(j)},
\quad A = 1,\dots,f.
\ee

The transition from the set
$\phi^{(j)}$
to
$\phi^{(A)}, \quad (\phi^{(A)})^{\dagger}$
is done with the invertible matrix $A$. Then one can easily prove that the
mass of the scalar field
$\phi^{(A)}$
is
$M_{A}$
and that the couple
$\phi^{(A)}, \quad f^{(A)}$
verifies the relations (\ref{wess}) corresponding to the mass
$M_{A}$.
$\qed$

It is interesting to note that, in some sense, the condition that the fields
are free is redundant. Indeed, suppose that only first order partial derivatives
can appear in the right hand side of (\ref{tensor}). Then the last
consistency relation (\ref{susy+lorentz}) is in our case:
\bea
\sum_{B=1}^{f} \left[ (\bar{\cal A}_{3} {\cal A}_{1})^{AB}
\sigma^{\mu}_{c\bar{b}} \partial_{\mu}f^{(B)}_{a}
+ ({\cal A}_{2} \bar{\cal A}_{1})^{AB} \epsilon_{ac} f^{(B)}_{\bar{b}} \right]
= 2 I_{f} \sigma^{\mu}_{a\bar{b}} \partial_{\mu}f^{(A)}_{c}.
\nonumber
\eea

For
$a = c$
we get
$\bar{\cal A}_{3} {\cal A}_{1} = 2 I_{f}$.
For
$a \not= c$
we obtain after contraction with
$\epsilon^{ac}$
\bea
- 2 \sigma^{\mu}_{a\bar{b}} \partial_{\mu}f^{(A)}_{a}
= \sum_{B=1}^{f} ({\cal A}_{2} \bar{\cal A}_{1})^{AB} f^{(B)}_{\bar{b}}
\nonumber
\eea
If we take into account that the fields are linearly independent, i.e.
no relations connect two different fields
$f^{(A)}$
then we conclude that the matrix
${\cal A}_{2} \bar{\cal A}_{1}$
should be diagonal i.e. we should have
${\cal A}_{2} \bar{\cal A}_{1} = - 2~i~M$
for some complex diagonal matrix $M$. The preceding equation becomes:
\bea
- i~\sigma^{\mu}_{a\bar{b}} \partial_{\mu}f^{(A)}_{a}
= M_{A} f^{(B)}_{\bar{b}}.
\nonumber
\eea
We can fix
$M_{A} \geq 0$
if we redefine the fields with a phase factor $\lambda$:
$f^{(A)} \mapsto \lambda~f^{(A)}$.
If we apply the operator
$- i~\sigma^{\mu}_{a\bar{b}} \partial_{\mu}$
to the last equation (\ref{ansatz}) then we easily obtain with the same
linear independence argument that the scalar fields should verify Klein-Gordon
equations.

Now we give the explicit expressions for the
 superfields associated to
the Wess-Zumino multiplet according to formul\ae~ (\ref{superfields})
and (\ref{wess}). We have:
\bea
\Phi(x,\theta,\bar{\theta})
\equiv W_{\theta,\bar{\theta}}~\phi(x)~W_{\theta,\bar{\theta}}^{-1}
\nonumber \\
= \phi - 2 \bar{\theta}^{\bar{a}} \bar{f}_{\bar{a}}
+ i~(\theta \sigma^{\mu} \bar{\theta}) \partial_{\mu}\phi
- m (\bar{\theta}\bar{\theta}) \phi^{\dagger}
- m (\bar{\theta}\bar{\theta}) \theta^{a} f_{a}
+ {m^{2}\over 4} (\bar{\theta}\bar{\theta}) (\theta\theta) \phi
\nonumber \\
F_{a}(x,\theta,\bar{\theta}) \equiv
W_{\theta,\bar{\theta}}~f_{a}(x)~W_{\theta,\bar{\theta}}^{-1}
\nonumber \\
= f_{a} - m \theta_{a} \phi
- i~\sigma^{\mu}_{a\bar{b}} \bar{\theta}^{\bar{b}} \partial_{\mu}\phi^{\dagger}
+ 2 m \theta_{a} \bar{\theta}^{\bar{b}} \bar{f}_{\bar{b}}
+ i~(\theta\sigma^{\mu}\bar{\theta}) \partial_{\mu}f_{a}
\nonumber \\
+ {i~m\over 2} (\theta\theta)
\sigma^{\mu}_{a\bar{b}} \bar{\theta}^{\bar{b}} \partial_{\mu}\phi
+ {m^{2}\over 2} (\bar{\theta}\bar{\theta}) \theta_{a} \phi^{\dagger}
- {m^{2}\over 4} (\bar{\theta}\bar{\theta}) (\theta\theta) f_{a}.
\label{scalar}
\eea
The result follows by computing the multiple commutators with the aid of
(\ref{wess}).

There are some very usefull relations
\bea
{\cal D}_{a} \Phi = 0, \qquad
{\cal D}_{a} \Phi^{\dagger} = 2 F_{a}
\nonumber \\
{\cal D}_{a} F_{b} = m~\epsilon_{ab} \Phi, \quad
{\cal D}_{a} \bar{F}_{\bar{b}} =
i\sigma_{a\bar{b}}^{\mu} \partial_{\mu}\Phi
\label{ddd}
\eea
which immediately follow from (\ref{d+s}). From the preceding relations one
can easily derive the following {\it super-equations of motions} for the WZ
superfield:
\bea
{\cal D}^{2} \Phi^{\dagger} = 4m \Phi
\qquad
\bar{\cal D}^{2} \Phi = 4m \Phi^{\dagger}.
\label{super-eq}
\eea

No variational principle needed in deriving these equations. The imaginary
unit in the ``sandwich formula" is essential!

We can obtain an interesting conclusion from (\ref{ddd}) and (\ref{super-eq})
concerning the triviality of the Lagrangian (\ref{phi3}); indeed we obtain
\be
:\Phi(X)^{3}: = {1 \over 4m} {\cal D}^{2}
\left[ :\Phi^{2}(X) \Phi(X)^{\dagger}: \right]
\ee
However the Wess-Zumino Lagrangian (\ref{wz-phi3}) is not trivial: one cannot
pull out the operator
${\cal D}^{2}$
in front of the $\delta$-factors.
\begin{rem}
Let us note that the first superfield (\ref{scalar}) seems to differ from the
usual Wess-Zumino (or chiral) superfield considered in the literature which
corresponds to a different choice of the operator (\ref{W-expo}) namely
without the imaginary factor in the exponential \cite{CS}. However, in
this case this operator is no longer unitary and the renormalizability
arguments from the previous Section should be reconsidered.

\end{rem}

Now the commutation relations can be obtained by direct computation; they are:
\bea
\left[ \Phi(x_{1},\theta_{1},\bar{\theta}_{1}),
\Phi(x_{2},\theta_{2},\bar{\theta}_{2}) \right]
\nonumber \\
=  2~i~m~
(\bar{\theta}_{1} - \bar{\theta}_{2}) (\bar{\theta}_{1} - \bar{\theta}_{2})
\exp[ i~(\theta_{1} \sigma^{\mu} \bar{\theta}_{1}
- \theta_{2} \sigma^{\mu} \bar{\theta}_{2}) \partial_{\mu} ]
D_{m}(x_{1} - x_{2} )
\nonumber \\
~\left[ \Phi(x_{1},\theta_{1},\bar{\theta}_{1}),
\Phi^{\dagger}(x_{2},\theta_{2},\bar{\theta}_{2}) \right]
\nonumber \\
= - 2~i~\exp[ i~(\theta_{1} \sigma^{\mu} \bar{\theta}_{1}
+ \theta_{2} \sigma^{\mu} \bar{\theta}_{2}
- 2~\theta_{2} \sigma^{\mu} \bar{\theta}_{1} ) \partial_{\mu} ]
D_{m}(x_{1} - x_{2} );
\label{commutators}
\eea
the commutation relations for the superfields
$F_{a}$
can be obtained from the preceding ones without explicit computations
from (\ref{ddd}). It has been noted in \cite{CS} that the super-order of
singularities are better than the general formula
 (\ref{super-order}),
namely:
\bea
\omega(D_{\Phi,\Phi^{\dagger}}) = -2, \quad \omega(D_{\Phi,\Phi}) = -3, \quad
\omega(D_{F_{a},F_{b}^{\dagger}}) = -1, \quad \omega(D_{F_{a}F_{b}}) = -2.
\eea
The relations (\ref{commutators}) explicitly show that the superfields
are causally 
 commuting, which is an essential ingredient to the
perturbative approach as 
 pointed out in Subsection \ref{axioms}.

We close by mentioning that superfields of the type (\ref{scalar}) will
play various 
 r\^oles in our supersymmetric extension of gauge
theories: we will need
 super-Fermi ghosts, super-Bose Higgs and
super-Bose for the matter field.

\subsection{The Ghost and Anti-Ghost Multiplets}

To construct a supersymmetric gauge theory it seems natural to extend in
a consistent super-symmetric way the usual ghost and anti-ghost fields.
It is rewarding that the preceding analysis goes through practically unchanged.
One only has to take care to invert the statistics assignment: the scalar
fields
$u^{(j)}, \quad j = 1,\dots,s'$
will be Hermitian and will respect Fermi-Dirac statistics; their Majorana
partners
$\chi^{(A)}_{a}, A = 1,\dots,f'$
will be bosons. This is enforced by the consistency relations
(\ref{susy+gauge1}).

The corresponding anti-ghost multiplet is denoted similarly: the scalar
fields
$\tilde{u}^{(j)}, \quad j = 1,\dots,s"$
will be anti-Hermitian and will respect Fermi-Dirac statistics; their
anti-Majorana partners
$\tilde{\chi}^{(A)}_{a}, A = 1,\dots,f"$
will be bosons:
\bea
(u^{(j)})^{\dagger} = u^{(j)}, \quad j = 1,\dots,s' \qquad
(\chi^{(A)}_{a})^{\dagger} = \bar{\chi}^{(A)}_{\bar{a}}, \quad A = 1,\dots,f'
\nonumber \\
(\tilde{u}^{(j)})^{\dagger} =  - \tilde{u}^{(j)}, \quad
j = 1,\dots,s^{\prime\prime} \qquad
(\tilde{\chi}^{(A)}_{a})^{\dagger} = - \bar{\tilde{\chi}}^{(A)}_{\bar{a}},
\quad A = 1,\dots,f^{\prime\prime}.
\eea

These are free fields; we have as before:
\bea
~[ \partial^{2} + (m'_{j})^{2} ] u^{(j)} = 0, \quad j = 1,\dots,s'
\nonumber \\
i~\sigma^{\mu}_{a\bar{b}} \partial_{\mu} \bar{\chi}^{(A)\bar{b}}
= M'_{A} \chi^{(A)}_{a}, \qquad
- i~\sigma^{\mu}_{a\bar{b}} \partial_{\mu} \chi^{(A)a}
= M'_{A} \bar{\chi}^{(A)}_{\bar{b}}, \quad A = 1,\dots,f'
\nonumber \\
~[ \partial^{2} + (m^{\prime\prime}_{j})^{2} ] \tilde{u}^{(j)} = 0, \quad
j = 1,\dots,s^{\prime\prime}
\nonumber \\
i~\sigma^{\mu}_{a\bar{b}} \partial_{\mu} \bar{\tilde{\chi}}^{(A)\bar{b}}
= M^{\prime\prime}_{A} \tilde{\chi}^{(A)}_{a}, \qquad
- i~\sigma^{\mu}_{a\bar{b}} \partial_{\mu} \tilde{\chi}^{(A)a}
= M^{\prime\prime}_{A} \bar{\tilde{\chi}}^{(A)}_{\bar{b}}, \quad
A = 1,\dots,f^{\prime\prime}.
\eea
The mass matrices are then
$m', m^{\prime\prime}, M', M^{\prime\prime}$.
The commutation relation involves a subtilty and will be dealt with
 later.

Now the changes in the argument of the preceding Subsection are minimal.
The unicity result of \cite{LW} stays true: one simply has to modify the
ansatz (\ref{ansatz}) to:
\bea
~\{ Q_{a}, u^{(j)} \} = \sum_{A=1}^{f} {\cal A}_{1}^{jA} \chi^{(A)}_{a}
\nonumber \\
i~[ Q_{a}, \chi^{(A)}_{b} ] = \epsilon_{ab} \sum_{j=1}^{s} {\cal A}_{2}^{Aj}
u^{(j)}
\qquad
i~[ Q_{a}, \bar{\chi}^{(A)}_{\bar{b}} ] = \sigma^{\mu}_{a\bar{b}}
\sum_{j=1}^{s} {\cal A}_{3}^{Aj} \partial_{\mu}u^{(j)}
\label{ansatz-gh}
\eea
and similarly for the anti-ghost multiplet. It is not very hard to see that
(\ref{susy+lorentz}) and (\ref{factor}) give again (\ref{wz1}) and (\ref{wz2}).
As a result we conclude that the ghost multiplet is a sum of elementary
ghosts multiplets built from a complex scalar field $u$ with Fermi statistics
and a Majorana spinor
$\chi$
with Bose statistics of the same mass
$m'$
such that we have instead of (\ref{wess}) the following
 relations:
\bea
~\{ Q_{a}, u \} = 0, \qquad \{ \bar{Q}_{\bar{a}}, u^{\dagger} \} = 0
\nonumber \\
~\{ Q_{a}, u^{\dagger} \} = 2 \chi_{a}, \qquad
~\{\bar{Q}_{\bar{a}}, u \} = 2 \bar{\chi}_{\bar{a}}
\nonumber \\
~[ Q_{a}, \chi_{b} ] = - m'~\epsilon_{ab} u, \qquad
~[ \bar{Q}_{\bar{a}}, \bar{\chi}_{\bar{b}} ]
= m'~\epsilon_{\bar{a}\bar{b}} u^{\dagger}
\nonumber \\
i~[ Q_{a}, \bar{\chi}_{\bar{b}} ] = \sigma^{\mu}_{a\bar{b}} \partial_{\mu}u
\qquad
i~[ \bar{Q}_{\bar{a}}, \chi_{b} \} = \sigma^{\mu}_{b\bar{a}}
\partial_{\mu}u^{\dagger}
\label{wess-gh}
\eea

For the anti-ghost multiplet we have instead:
\bea
~\{ Q_{a}, \tilde{u} \} = 0, \qquad
\{ \bar{Q}_{\bar{a}}, \tilde{u}^{\dagger} \} = 0
\nonumber \\
~\{ Q_{a}, \tilde{u}^{\dagger} \} = 2 \tilde{\chi}_{a}, \qquad
~\{\bar{Q}_{\bar{a}}, \tilde{u} \} = - 2 \bar{\tilde{\chi}}_{\bar{a}}
\nonumber \\
~[ Q_{a}, \tilde{\chi}_{b} ] = m^{\prime\prime}~\epsilon_{ab} \tilde{u}, \qquad
~[ \bar{Q}_{\bar{a}}, \bar{\tilde{\chi}}_{\bar{b}} ]
= m^{\prime\prime}~\epsilon_{\bar{a}\bar{b}} \tilde{u}^{\dagger}
\nonumber \\
i~[ Q_{a}, \bar{\tilde{\chi}}_{\bar{b}} ]
= - \sigma^{\mu}_{a\bar{b}} \partial_{\mu}\tilde{u}
\qquad
i~[ \bar{Q}_{\bar{a}}, \tilde{\chi}_{b} \} = \sigma^{\mu}_{b\bar{a}}
\partial_{\mu}\tilde{u}^{\dagger}
\label{wess-antigh}
\eea
where the changes of some signs follows from the different behaviour of
$\tilde{\chi}$
with 
respect to Hermitian conjugation.

One can easily prove that all consistency conditions (\ref{lorentz}),
(\ref{susy+lorentz}), (\ref{factor})  are verified. We call these multiplets
the {\it ghost} (resp.) {\it anti-ghost multiplets}.

To consider the commutation relations we remember that for 
 usual gauge
theories \cite{Sc} one has to consider that the ghost and the
anti-ghost fields are of the same mass and verify commutation relations
of
 the following type:
\bea
\left\{ u_{j}(x), \tilde{u}_{k}(y) \right\} = i~\delta_{jk} D_{m_{j}}(x-y).
\nonumber
\eea

It is natural to postulate
\be
~\left\{ u(x), \tilde{u}^{\dagger}(y) \right\} = 2i~ D_{m'}(x-y)
;
\label{CCR-wess-gh1}
\ee
then we have from (\ref{susy+CCR}) 
$m' = m^{\prime\prime}$
 and
\bea
~[ \chi_{a}(x), \tilde{\chi}_{b}(y) ] = - i~m'~\epsilon_{ab}~D_{m'}(x-y),
\nonumber \\
~[ \bar{\chi}_{\bar{a}}(x), \tilde{\chi}_{b}(y) ]
= - \sigma^{\mu}_{b\bar{a}}~\partial_{\mu}D_{m'}(x-y);
\label{CCR-wess-gh2}
\eea
all other (anti)commutators are zero. A further check of consistency we will
get when the gauge charge $Q$ will be introduced.

We now give the explicit expressions for the superfields associated to the
ghost and anti-ghost multiplets using the formula  (\ref{superfields}) and
(\ref{wess-gh}) + (\ref{wess-antigh}). We have for the ghost multiplet:
\bea
U(x,\theta,\bar{\theta})
\equiv W_{\theta,\bar{\theta}}~u(x)~W_{\theta,\bar{\theta}}^{-1}
\nonumber \\
= u - 2~i~ \bar{\theta}^{\bar{a}} \bar{\chi}_{\bar{a}}
+ i~(\theta \sigma^{\mu} \bar{\theta}) \partial_{\mu}u
- m' (\bar{\theta}\bar{\theta}) u^{\dagger}
- i~m' (\bar{\theta}\bar{\theta}) \theta^{a} \chi_{a}
+ {m'^{2}\over 4} (\bar{\theta}\bar{\theta}) (\theta\theta) u
\nonumber \\
X_{a}(x,\theta,\bar{\theta}) \equiv
W_{\theta,\bar{\theta}}~\chi_{a}(x)~W_{\theta,\bar{\theta}}^{-1}
\nonumber \\
= \chi_{a} + i~ m' \theta_{a} u
- \sigma^{\mu}_{a\bar{b}} \bar{\theta}^{\bar{b}} \partial_{\mu}u^{\dagger}
+ 2 m' \theta_{a} \bar{\theta}^{\bar{b}} \bar{\chi}_{\bar{b}}
+ i~(\theta\sigma^{\mu}\bar{\theta}) \partial_{\mu}\chi_{a}
\nonumber \\
+ {m'\over 2} (\theta\theta)
\sigma^{\mu}_{a\bar{b}} \bar{\theta}^{\bar{b}} \partial_{\mu}u
- {i~m'^{2}\over 2} (\bar{\theta}\bar{\theta}) \theta_{a} u^{\dagger}
- {m'^{2}\over 4} (\bar{\theta}\bar{\theta}) (\theta\theta) \chi_{a}
\eea
and respectively for the anti-ghost multiplet:
\bea
\tilde{U}(x,\theta,\bar{\theta})
\equiv W_{\theta,\bar{\theta}}~\tilde{u}(x)~W_{\theta,\bar{\theta}}^{-1}
\nonumber \\
= \tilde{u} + 2~i~ \bar{\theta}^{\bar{a}} \bar{\tilde{\chi}}_{\bar{a}}
+ i~(\theta \sigma^{\mu} \bar{\theta}) \partial_{\mu}\tilde{u}
+ m' (\bar{\theta}\bar{\theta}) \tilde{u}^{\dagger}
+ i~m' (\bar{\theta}\bar{\theta}) \theta^{a} \tilde{\chi}_{a}
+ {m'^{2}\over 4} (\bar{\theta}\bar{\theta}) (\theta\theta) \tilde{u}
\nonumber \\
\tilde{X}_{a}(x,\theta,\bar{\theta}) \equiv
W_{\theta,\bar{\theta}}~\tilde{\chi}_{a}(x)~W_{\theta,\bar{\theta}}^{-1}
\nonumber \\
= \tilde{\chi}_{a} - i~ m' \theta_{a} \tilde{u}
- \sigma^{\mu}_{a\bar{b}} \bar{\theta}^{\bar{b}}
\partial_{\mu}\tilde{u}^{\dagger}
+ 2 m' \theta_{a} \bar{\theta}^{\bar{b}} \bar{\tilde{\chi}}_{\bar{b}}
+ i~(\theta\sigma^{\mu}\bar{\theta}) \partial_{\mu}\tilde{\chi}_{a}
\nonumber \\
- {m'\over 2} (\theta\theta)
\sigma^{\mu}_{a\bar{b}} \bar{\theta}^{\bar{b}} \partial_{\mu}\tilde{u}
- {i~m'^{2}\over 2} (\bar{\theta}\bar{\theta}) \theta_{a} \tilde{u}^{\dagger}
- {m'^{2}\over 4} (\bar{\theta}\bar{\theta}) (\theta\theta) \tilde{\chi}_{a}.
\eea

As for the Wess-Zumino multiplet we have from (\ref{s}) equations of the
type (\ref{ddd}):
\bea
{\cal D}_{a} U = 0, \qquad
{\cal D}_{a} U^{\dagger} = 2~i~X_{a}
\nonumber \\
{\cal D}_{a} X_{b} = - i~m'~\epsilon_{ab} U, \quad
{\cal D}_{a} \bar{X}_{\bar{b}} =
\sigma_{a\bar{b}}^{\mu} \partial_{\mu}U
\label{ddd-gh}
\eea
and similarly for the anti-ghosts. Super-equations of motion follow:
\bea
{\cal D}^{2} U^{\dagger} = 4m'~U
\qquad
\bar{\cal D}^{2} U = 4m'~U^{\dagger}
\nonumber \\
{\cal D}^{2} \tilde{U}^{\dagger} = 4m'~\tilde{U}
\qquad
\bar{\cal D}^{2} \tilde{U} = 4m'~\tilde{U}^{\dagger}.
\label{super-eq-gh}
\eea

Now the corresponding commutation relations are:
\bea
~\left\{ U(x_{1},\theta_{1},\bar{\theta}_{1}),
\tilde{U}(x_{2},\theta_{2},\bar{\theta}_{2}) \right\}
\nonumber \\
=  2~i~m'~
(\bar{\theta}_{1} - \bar{\theta}_{2}) (\bar{\theta}_{1} - \bar{\theta}_{2})
\exp[ i~(\theta_{1} \sigma^{\mu} \bar{\theta}_{1}
- \theta_{2} \sigma^{\mu} \bar{\theta}_{2}) \partial_{\mu} ]
D_{m'}(x_{1} - x_{2} )
\nonumber \\
~\left\{ U(x_{1},\theta_{1},\bar{\theta}_{1}),
\tilde{U}^{\dagger}(x_{2},\theta_{2},\bar{\theta}_{2}) \right\}
\nonumber \\
= 2~i~\exp[ i~(\theta_{1} \sigma^{\mu} \bar{\theta}_{1}
+ \theta_{2} \sigma^{\mu} \bar{\theta}_{2}
- 2~\theta_{2} \sigma^{\mu} \bar{\theta}_{1} ) \partial_{\mu} ]
D_{m'}(x_{1} - x_{2} );
\eea
the commutation relations for the superfields
$X_{a}$
can be obtained from the preceding ones without explicit computations from
(\ref{ddd-gh}). The super-order of singularities are better that the general
formula
 (\ref{super-order}), namely:
\bea
\omega(D_{U,\tilde{U}^{\dagger}}) = -2, \quad \omega(D_{U,\tilde{U}}) = -3.
\eea

\subsection{The Vector Multiplet}

To construct a gauge theory one needs a multiplet including a spin $1$
field. To obtain such multiplets is not so easy as in the Wess-Zumino
case. The usual vector multiplet from the literature \cite{WB},
\cite{Wes} contains scalar, Majorana and vector component fields.
Detailed checks of the consistency relations outlined in Subsection
\ref{susy-multiplets} seems to be absent from the literature. We will
consider here a new vector multiplet which has the nice property that
the corresponding gauge structure is similar to the usual gauge
theories. If this model is consistent with the phenomenology
 it brings
new physics, as will be seen in the next Section.

First we should clear up why unicity theorems of the type presented above
in Subsection \ref{wz-multiplet} are not available. This point is also
emerging from the analysis of \cite{LW}. Let us consider first the next
possible generalization of the Wess-Zumino scheme. We take as basic fields
some vector fields
$v_{\mu}^{(j)}, \quad j = 1,\dots,v$
with Bose statistics and some
Majorana fields
$\psi^{(A)}_{a}, A = 1,\dots,f$
with Fermi statistics. Equations of motion of the type (\ref{kg+dirac}) are
also assumed. As we have said in Subsection  \ref{axioms} we do not impose
the transversality condition (\ref{transv}) in order to have good singularity
behaviour of the causal functions. (In \cite{LW} a more general situation is
considered, i.e. one considers some scalar fields also but the transversality
condition is imposed).

Now we can write the most general ansatz of the type (\ref{ansatz}) for this
case; it is:
\bea
i~[Q_{a}, v_{\mu}^{(j)} ] = \sum_{A=1}^{f} \left[
{\cal A}_{1}^{jA} \partial^{\mu} \psi^{(A)}_{a}
+ {\cal A}_{4}^{jA} \sigma^{\mu}_{a\bar{b}} \bar{\psi}^{(A)\bar{b}}
+ {\cal A}_{5}^{jA} \sigma^{\mu\nu}_{ab} \partial_{\nu} \psi^{(A)b} \right]
\nonumber \\
\{ Q_{a}, \psi^{(A)}_{b} \} = \sum_{j=1}^{s} \left[
\epsilon_{ab} {\cal A}_{2}^{Aj} \partial^{\mu}v_{\mu}^{(j)}
+ {\cal A}_{6}^{jA} \sigma^{\mu\nu}_{ab} v^{(j)}_{\mu\nu} \right]
\nonumber \\
\{ Q_{a}, \bar{\psi}^{(A)}_{\bar{b}} \} = \sigma^{\mu}_{a\bar{b}}
\sum_{j=1}^{v} {\cal A}_{3}^{Aj} v_{\mu}^{(j)}
\label{v+m}
\eea
where we use the well-known notations
\be
v^{(j)}_{\mu\nu} \equiv \partial_{\mu} v^{(j)}_{\nu}
- (\mu \leftrightarrow \nu)
\ee
and
\be
\sigma^{\mu\nu}_{ab} \equiv {i\over 4}
\left[ \sigma^{\mu}_{a\bar{c}} \epsilon^{\bar{c}\bar{d}}
\sigma^{\nu}_{b\bar{d}} - (\mu \leftrightarrow \nu) \right].
\ee

One can proceed as in Subsection \ref{wz-multiplet} and write down all the
relations following from the consistency conditions, but as in \cite{LW}, a
general solution seems to be impossible to obtain: there are ``too many"
matrices
${\cal A}_{i}$!
Another possibility is to construct the multiplet directly from the
Jacobi consistency conditions. This method is used in \cite{CGS}. If we
start with one Majorana field
$f = 1$,
it turns out that we need two Hermitian vector fields
$v = 2$,
{\bf but in addition a spin} $3/2$ {\bf field}. The above problem for
$f = 1$
has no solution.

This leads us consider a related situation in which we replace the above
spinor fields by some Majorana-Rarita-Schwinger fields
$\psi^{(A)}_{\mu a}, A = 1,\dots,f$
(without using any spinor field).
First we fix all conditions on the free fields of the model. We require
:
\begin{itemize}
\item
The fields
$\psi^{(A)}_{\mu a}, A = 1,\dots,f$
behave as spinors with respect to the index $a$ and as vectors with respect
to the index $\mu$.
\item
Hermiticity:
\be
(v_{\mu}^{(j)})^{\dagger} = v_{\mu}^{(j)}, \quad j = 1,\dots,v \qquad
(\psi^{(A)}_{\mu a})^{\dagger} = \bar{\psi}^{(A)}_{\mu\bar{a}},
\quad A = 1,\dots,f.
\ee
\item
Equations of motion:
The fields verify Klein-Gordon (resp. Dirac) equation:
\bea
(\partial^{2} + m_{j}^{2}) v_{\mu}^{(j)} = 0, \quad j = 1,\dots,s
\nonumber \\
i~\sigma^{\nu}_{a\bar{b}} \partial_{\nu} \bar{\psi}^{(A)\bar{b}}_{\mu}
= M_{A} \psi^{(A)}_{\mu a}, \qquad
- i~\sigma^{\nu}_{a\bar{b}} \partial_{\nu} \psi^{(A)a}_{\mu}
= M_{A} \bar{\psi}^{(A)}_{\mu\bar{b}}, \quad A = 1,\dots,f.
\label{kg+dirac-v}
\eea
\item
(Anti)commutation relations:
\bea
\left[ v_{\mu}^{(j)}(x), v_{\rho}^{(k)}(y)^{\dagger} \right]
= 2~i~\delta_{jk} g_{\mu\rho}~ D_{m_{j}}(x-y),
\nonumber \\
\left\{ \psi^{A}_{\mu a}(x), \psi^{B}_{\rho b}(y) \right\}
= -~i~\delta_{AB}~g_{\mu\rho}~M_{A}~ \epsilon_{ab}~D_{M_{A}}(x-y),
\nonumber \\
\left\{ \psi^{A}_{\mu a}(x), \bar{\psi}^{B}_{\rho\bar{b}}(y) \right\} =
- \delta_{AB}~g_{\mu\rho} \sigma^{\mu}_{a\bar{b}}~\partial_{\mu}D_{M_{A}}(x-y),
\nonumber \\
\left[ v_{\mu}^{(j)}(x), \psi^{(A)}_{\rho a}(y) \right] = 0.
\label{CCR-v}
\eea
\end{itemize}
We do not impose a transversality condition of the type (\ref{transv}) for
the same reason as explained before.

The analogue of (\ref{v+m}) is now:
\bea
i~[Q_{a}, v_{\mu}^{(j)} ] = \sum_{A=1}^{f} \left[
{\cal A}_{1}^{jA} \psi^{(A)}_{\mu a}
+ {\cal A}_{4}^{jA} (\sigma_{\mu})_{a\bar{b}}
\partial^{\nu}\bar{\psi}^{(A)\bar{b}}_{\nu}
+ {\cal A}_{5}^{jA} \sigma^{\nu}_{a\bar{b}}
\partial_{\mu}\bar{\psi}^{(A)\bar{b}}_{\nu}
+ {\cal A}_{6}^{jA} (\sigma_{\mu\nu})_{ab} \psi^{(A)\nu b}
\right]
\nonumber \\
\{ Q_{a}, \psi^{(A)}_{\mu b} \} = \sum_{j=1}^{s} \left[
\epsilon_{ab} {\cal A}_{2}^{Aj} v_{\mu}^{(j)}
+ {\cal A}_{7}^{jA} (\sigma_{\mu\nu})_{ab} v^{(j)\nu} \right]
\nonumber \\
\{ Q_{a}, \bar{\psi}^{(A)}_{\mu\bar{b}} \} = \sum_{j=1}^{v} \left[
{\cal A}_{3}^{Aj} \sigma^{\nu}_{a\bar{b}} \partial_{\nu}v_{\mu}^{(j)}
+ {\cal A}_{8}^{Aj} \sigma^{\nu}_{a\bar{b}} \partial_{\mu}v_{\nu}^{(j)} \right].
\label{v+rs}
\eea

The number of undetermined matrices proliferates. However there is a
particular case when the problem can be analysed completely, namely when
we have:
${\cal A}_{i} = 0, \quad i = 6,7,8$
i.e.
the preceding ansatz takes the form:
\bea
i~[Q_{a}, v_{\mu}^{(j)} ] = \sum_{A=1}^{f} \left[
{\cal A}_{1}^{jA} \psi^{(A)}_{\mu a}
+ {\cal A}_{4}^{jA} (\sigma_{\mu})_{a\bar{b}}
\partial^{\nu}\bar{\psi}^{(A)\bar{b}}_{\nu}
+ {\cal A}_{5}^{jA} \sigma^{\nu}_{a\bar{b}}
\partial_{\mu}\bar{\psi}^{(A)\bar{b}}_{\nu} \right]
\nonumber \\
\{ Q_{a}, \psi^{(A)}_{\mu b} \} = \sum_{j=1}^{s}
\epsilon_{ab} {\cal A}_{2}^{Aj} v_{\mu}^{(j)}
\qquad
\{ Q_{a}, \bar{\psi}^{(A)}_{\mu\bar{b}} \} = \sum_{j=1}^{v}
{\cal A}_{3}^{Aj} \sigma^{\nu}_{a\bar{b}} \partial_{\nu}v_{\mu}^{(j)}.
\label{v+rs-red}
\eea

In this case the consistency relations are not very complicated:
\bea
{\cal A}_{i} {\cal A}_{3} = 0, \quad i = 4, 5 \qquad
{\cal A}_{i} {\cal A}_{j} = 0, \quad i = 2,3 \quad j = 1, 4, 5
\nonumber \\
\bar{\cal A}_{1} {\cal A}_{3} + {\cal A}_{1} \bar{\cal A}_{3} = 2 I_{s},
\qquad \bar{\cal A}_{3} {\cal A}_{1} = 2 I_{f}
\nonumber \\
\bar{\cal A}_{i} {\cal A}_{2} + {\cal A}_{i} \bar{\cal A}_{2} = 0,
\quad i = 4, 5
\nonumber \\
\bar{\cal A}_{i} {\cal A}_{j} = 0, \quad i = 2, 3 \quad j = 4, 5
\nonumber \\
{\cal A}_{2} \bar{\cal A}_{1} = - 2~i~M
\label{v1}
\eea
and
\bea
{\cal A}_{2} = -~i~M~{\cal A}_{3}, \qquad
M~{\cal A}_{2} = -~i~{\cal A}_{3}~m^{2}, \qquad
m^{2}~{\cal A}_{i} = {\cal A}_{i}~M^{2}, \quad i= 1, 3, 4, 5.
\label{v2}
\eea

As in Subsection \ref{wz-multiplet} we get
$s = 2f$.
Next, we define the matrices
$A, B \in M_{\R}(s,s)$
as in (\ref{AB}) and find out that they are inverse to each other (up to a
factor $2$). Finally we replace the $s$ real vector fields
$v_{\mu}^{(j)}$
by
$f = s/2$
complex fields according to:
\be
v_{\mu}^{(A)} \equiv \sum_{j=1}^{s} {\cal A}^{Aj}_{3} v_{\mu}^{(j)},
\quad A = 1,\dots,f
\ee
and the multiplet decouples into a sum of new vector multiplets where by
definition such a multiplet is built from a complex vector field
$v_{\mu}$
and a Majorana-Rarita-Schwinger field
$\psi_{\mu a}$
subject to the following consistency conditions:
\begin{itemize}
\item
Hermiticity
\be
(\psi_{\mu a})^{\dagger} = \bar{\psi}_{\mu\bar{a}}.
\ee
\item
Equations of motion:
The fields verify Klein-Gordon (resp. Dirac) equation with the same mass
\bea
(\partial^{2} + M^{2}) v_{\mu} = 0,
\nonumber \\
i~\sigma^{\nu}_{a\bar{b}} \partial_{\nu} \bar{\psi}^{\bar{b}}_{\mu}
= M \psi_{\mu a}, \qquad
- i~\sigma^{\nu}_{a\bar{b}} \partial_{\nu} \psi^{a}_{\mu}
= M \bar{\psi}_{\mu\bar{b}}.
\label{kg+dirac-v-red}
\eea
\item
(Anti)commutation relations:
\bea
~\left[ v_{\mu}(x), v_{\rho}^{\dagger}(y) \right]
= 2~i~g_{\mu\rho}~ D_{M}(x-y),
\nonumber \\
~\left\{ \psi_{\mu a}(x), \psi_{\rho b}(y) \right\}
= -~i~M~g_{\mu\rho} \epsilon_{ab}~D_{M}(x-y),
\nonumber \\
~\left\{ \psi_{\mu a}(x), \bar{\psi}_{\rho\bar{b}}(y) \right\} =
- g_{\mu\rho} \sigma^{\mu}_{a\bar{b}}~\partial_{\mu}D_{M}(x-y),
\nonumber \\
~\left[ v_{\mu}(x), \psi_{\rho a}(y) \right] = 0.
\label{CCR-v-red}
\eea
\item
The action of the supercharges:
\bea
~[Q_{a}, v_{\mu} ] = 0, \qquad [\bar{Q}_{\bar{a}}, v^{\dagger}_{\mu} ] = 0
\nonumber \\
i~[Q_{a}, v_{\mu}^{\dagger} ] = 2~\psi_{\mu a}, \qquad
i~[\bar{Q}_{\bar{a}}, v_{\mu} ] = 2~\bar{\psi}_{\mu\bar{a}}
\nonumber \\
~\{ Q_{a}, \psi_{\mu b} \} = - i~\epsilon_{ab} M v_{\mu}, \qquad
\{ \bar{Q}_{\bar{a}}, \bar{\psi}_{\mu\bar{b}} \}
= i~\epsilon_{\bar{a}\bar{b}} M v_{\mu}^{\dagger},
\nonumber \\
~\{ Q_{a}, \bar{\psi}_{\mu\bar{b}} \} = \sigma^{\nu}_{a\bar{b}}
\partial_{\nu}v_{\mu}, \qquad
\{ \bar{Q}_{\bar{a}}, \psi_{\mu b} \} = \sigma^{\nu}_{b\bar{a}}
\partial_{\nu}v_{\mu}^{\dagger}.
\label{susy-v}
\eea
\end{itemize}

We call this new multiplet {\it the RS vector multiplet}.

The associated superfield can be easily constructed in analogy to the
case studied in Subsection \ref{wz-multiplet}. We have:
\bea
V_{\mu}(x,\theta,\bar{\theta})
\equiv W_{\theta,\bar{\theta}}~v_{\mu}(x)~W_{\theta,\bar{\theta}}^{-1}
\nonumber \\
= v_{\mu} - 2 \bar{\theta}^{\bar{a}} \bar{\psi}_{\mu\bar{a}}
+ i~(\theta \sigma^{\nu} \bar{\theta}) \partial_{\nu}v_{\mu}
- M (\bar{\theta}\bar{\theta}) v_{\mu}^{\dagger}
- M (\bar{\theta}\bar{\theta}) \theta^{a} \psi_{\mu a}
+ {M^{2}\over 4} (\bar{\theta}\bar{\theta}) (\theta\theta) v_{\mu}
\nonumber \\
\Psi_{\mu a}(x,\theta,\bar{\theta}) \equiv
W_{\theta,\bar{\theta}}~\psi_{\mu a}(x)~W_{\theta,\bar{\theta}}^{-1}
\nonumber \\
= \psi_{\mu a} - M \theta_{a} v_{\mu}
- i~\sigma^{\nu}_{a\bar{b}} \bar{\theta}^{\bar{b}} \partial_{\nu}v_{\mu}
+ 2 M \theta_{a} \bar{\theta}^{\bar{b}} \bar{\psi}_{\mu\bar{b}}
+ i~(\theta\sigma^{\nu}\bar{\theta}) \partial_{\nu}\psi_{\mu a}
\nonumber \\
+ {i~M\over 2} (\theta\theta)
\sigma^{\nu}_{a\bar{b}} \bar{\theta}^{\bar{b}} \partial_{\nu}v_{\mu}
+ {M^{2}\over 2} (\bar{\theta}\bar{\theta}) \theta_{a} v_{\mu}
- {M^{2}\over 4} (\bar{\theta}\bar{\theta}) (\theta\theta) \psi_{\mu a}
\label{RS}
\eea
and the commutation relations are:
\bea
\left[ V_{\mu}(x_{1},\theta_{1},\bar{\theta}_{1}),
V_{\rho}(x_{2},\theta_{2},\bar{\theta}_{2}) \right]
\nonumber \\
=  - 2~i~M~g_{\mu\rho}
(\bar{\theta}_{1} - \bar{\theta}_{2}) (\bar{\theta}_{1} - \bar{\theta}_{2})
\exp[ i~(\theta_{1} \sigma^{\nu} \bar{\theta}_{1}
- \theta_{2} \sigma^{\nu} \bar{\theta}_{2}) \partial_{\nu} ]
D_{M}(x_{1} - x_{2} )
\nonumber \\
~\left[ V_{\mu}(x_{1},\theta_{1},\bar{\theta}_{1}),
V_{\rho}^{\dagger}(x_{2},\theta_{2},\bar{\theta}_{2}) \right]
\nonumber \\
= 2~i~g_{\mu\rho} \exp[ i~(\theta_{1} \sigma^{\nu} \bar{\theta}_{1}
+ \theta_{2} \sigma^{\nu} \bar{\theta}_{2}
- 2~\theta_{2} \sigma^{\nu} \bar{\theta}_{1} ) \partial_{\nu} ]
D_{M}(x_{1} - x_{2} ).
\label{ccr-mrs}
\eea

\subsection{The Gauge Supermultiplet}

According to the usual wisdom of ordinary quantum gauge theory, one has to
``gauge away" the unphysical degrees
 of freedom of a vector field using ghost
fields. For a vector field
$v_{\mu}$
of positive mass $m$ one associates to it three fields
$u,~\tilde{u},~\phi$
such that:
\begin{itemize}
\item
All three are scalar fields;
\item
All them have the same mass $m$ as the vector field.
\item
The Hermiticity properties are;
\be
\phi^{\dagger} = \phi, \qquad u^{\dagger} = u, \qquad
\tilde{u}^{\dagger} = - \tilde{u}
\ee
\item
The first two ones
$u, \quad \tilde{u}$
are fermionic and
$\phi$
is bosonic.
\item
The commutation relations are:
\bea
~[ \phi(x), \phi(y) ] = - i~D_{m}(x - y), \qquad
~\{ u(x), \tilde{u}(y) \} = - i~D_{m}(x - y)
\eea
and the rest of the (anti)commutators are zero.
\end{itemize}

Then one introduces the {\it gauge charge} $Q$ according to:

\bea
Q \Omega = 0, \qquad Q^{\dagger} = Q,
\nonumber \\
~[ Q, v_{\mu} ] = i \partial_{\mu}u, \qquad [ Q, \phi ] = i~m~u
\nonumber \\
~\{ Q, u \} = 0, \qquad
\{ Q, \tilde{u} \} = - i~(\partial^{\mu}v_{\mu} + m~\phi).
\label{gh-charge}
\eea

It can be proved that this gauge charge is well defined by these relations
i.e. it is compatible with the (anti)commutation relations.
Moreover one has
$Q^{2} = 0$
so the factor space
$Ker(Q)/Im(Q)$
makes sense; it can be proved that this is the physical space of an ensemble
of identical particles of spin $1$. For details see \cite{Sc}, \cite{Gr1}.

Now it makes sense to copy this structure for the superfields. We will
simply replace
\be
v_{\mu} \rightarrow V_{\mu}, \qquad
\phi \rightarrow \Phi, \qquad
u \rightarrow U, \qquad \tilde{u} \rightarrow \tilde{U}
\label{subst}
\ee
where all these multiplets are of the same positive mass $m$.
We will prove that the structure so obtained is consistent. In other word, we
try to {\it define} the supercharge $Q$ such that:
\bea
Q \Omega = 0,
\nonumber \\
~[ Q, V_{\mu} ] = i \partial_{\mu}U, \qquad
[ Q, \Phi ] = i~m~U
\nonumber \\
~\{ Q, U \} = 0, \qquad
\{ Q, \tilde{U} \} = - i~(\partial^{\mu}V_{\mu} + m \Phi).
\label{def-Q}
\eea

It is not at all obvious that all these relations are consistent. This would be
true if the Hilbert space
${\cal H}_{G}$
would be generated acting on the vacuum
$\Omega$
only with the superfields. But this is not true: the generic form of a vector
being (\ref{hilbert}).

It is a remarkable fact that the preceding construction is indeed consistent:
the preceding relations are equivalent with the following set of commutation
properties in terms of component fields:
\bea
~[ Q, v_{\mu} ] = i \partial_{\mu}u, \qquad
~[ Q, \psi_{\mu a} ] = \partial_{\mu}\chi_{a}, \qquad
\nonumber \\
~[ Q, \phi ] = i~m~u, \qquad
\{ Q, f_{a} \} = m~\chi_{a}
\nonumber \\
~\{ Q, u \} = 0, \qquad
[ Q, \chi_{a} ] = 0,
\nonumber \\
~\{ Q, \tilde{u} \} = - i~(\partial^{\mu}v_{\mu} + m~\phi), \qquad
[ Q, \tilde{\chi} ] = - (\partial^{\mu}\psi_{\mu a} + m~f_{a})
\eea
and the relations which follow from Hermitian conjugation.

The final check is to prove that the consistency relations (\ref{susy+gauge1})
are true and this easily follows. It also can be showed that, as for
the usual gauge case, the
factor space
$Ker(Q)/Im(Q)$
describes a system of identical
$\Omega_{1}$
super-symmetric systems. So, the analogy with the usual gauge case is really
remarkable.
 Moreover, it is quite easy to obtain consistent gauge invariant
couplings for this multiplet. For instance, we can generalize the
Abelian Higgs-Kibble model from \cite{Sc}, Section 4.1 as follows.
We define the following superfields:
\bea
V'_{\mu}(X) \equiv V_{\mu}(X) + V_{\mu}^{\dagger}(X), \quad
U'(X) \equiv U(X) + U^{\dagger}(X), \quad
\tilde{U}'(X) \equiv \tilde{U}(X) - \tilde{U}^{\dagger}(X), \quad
\nonumber \\
\Phi'(X) \equiv \Phi(X) + \Phi^{\dagger}(X), \quad
\Phi_{H}'(X) \equiv \Phi_{H}(X) + \Phi_{H}^{\dagger}(X)
\eea
where
$\Phi_{H}$
is a Wess-Zumino multiplet of mass
$m_{H}$
and we extend the definition of the gauge charge $Q$ admitting that, besides
(\ref{def-Q}) we also have:
\bea
[Q, \Phi_{H} ] = 0.
\label{gauge-H}
\eea
Then the action of the gauge charge on the new superfields verifies the
relations (\ref{def-Q}).
So we can substitute everywhere in the expression (4.1.19) from \cite{Sc}
the ordinary fields by ``primed"-superfields i.e. we have:
\bea
T(X) = m~V^{\prime\mu}V^{\prime}_{\mu} \Phi^{\prime}_{H} +
U^{\prime} \tilde{U}^{\prime} \Phi^{\prime}_{H}
- V^{\prime}_{\mu} (\Phi^{\prime}_{H} \partial^{\mu}\Phi^{\prime} -
\Phi^{\prime} \partial^{\mu}\Phi^{\prime}_{H})
- {m_{H}^{2} \over 2m} \left[\Phi^{\prime}_{H} (\Phi^{\prime})^{2}
+ (\Phi^{\prime})_{H}^{3} \right]
\eea

One can even add to this expression a interaction with ``matter" fields
of the type

\bea
F^{a}(X) \sigma^{\mu}_{a\bar{b}} \bar{F}^{\bar{b}}(X)~ V_{\mu}(X),
\eea
where the superfield
$F^{a}$
has the structure given by the second part of the formula (\ref{scalar}) for
some mass $M$ and verifies a condition similar to (\ref{gauge-H}). Gauge
invariance in the first order of the perturbation theory follows elementary.
Gauge invariance in the second and third order of perturbation theory is more
subtle: if we use theorem \ref{super} we can argue that if the ordinary gauge
model does not have anomalies in these orders, then the anomalies are absent
for the supersymmetric extension too.

In the next Section we will construct consistent self-interactions of
two superfields of the type
$V^{\mu}$
which will be the supersymmetric extensions of the four spin $1$ bosons
appearing in the electroweak theory: one photon and three heavy Bosons
but we will use explicit supersymmetry breaking such that the superpartners
of the Bosons
$W^{\pm}$
and $Z$ became very heavy.

We close by mentioning that in \cite{CS}, \cite{CGS} one can find
another vector multiplet containing, beside the vector and the
Majorana-Rarita-Schwinger fields, a supplementary spin $1/2$ field.
However, the latter is determined in terms of the MRS field so the two
multiplets are in fact related.

\section{Supersymmetric Extension of the Electroweak Theory\label{ew}}

Another remarkable fact connected with our gauge construction is the fact that
one can take the unicity results concerning the interaction from \cite{Sc} and
\cite{Gr1} as they are and only make the substitution (\ref{subst}). In
this way the number of free parameters of the supersymmetric extension
of the standard model does not increases as in the usual approaches based
on the usual vector multiplet. According to the analysis from Subsection
\ref{axioms} the renormalizability of this model is saved in spite of the
fact that expressed in components terms of canonical dimension $6$
seems to spoil this property. We also mention that the gauge anomalies
cannot be eliminated using supersymmetry invariance.

To consider a concrete supergauge theory we cannot turn to supersymmetric
QED because it has one Hermitian gauge field only, the photon. But in our
supersymmetric extension we work with complex gauge fields, therefore we
need an even number of Hermitian ones. Instead of studying some
theoretical model we want directly investigate the electroweak theory.
Here the two $W^\pm$-bosons naturally belong to one complex supergauge
field $W^\mu (x,\theta,\bar\theta)$,
$$W^\mu=(W_1^\mu-iW_2^\mu)/\sqrt{2},\quad W^{\mu\dagger}=(W_1^\mu+
iW_2^\mu)/\sqrt{2},\eqno(4.1)$$
its spin-3/2 components are assumed to be heavy due to breaking of
supersymmetry. A simple possibility to achieve the breaking is by adding
 a quadratic interaction term
$\lambda g_1(x)(\psi_\mu^a\psi^\mu_a+\bar\psi_{\bar a}^\mu\bar\psi_\mu^{\bar
a})$. Such an interaction can be resummed to all orders in $\lambda$ in
the adiabatic limit $g_1(x)\to 1$ and results in a mass change of the
$\psi^\mu_a$-field. The $Z$-boson
and the photon are members of a second supergauge field
$$V^\mu=(Z^\mu-iA^\mu)/\sqrt{2},\quad V^{\mu\dagger}=(Z^\mu+
iA^\mu)/\sqrt{2}.\eqno(4.2)$$
Here the susy-breaking is even stronger because the $Z$ and the photon
have different mass.

In contrast to supersymmetry, gauge invariance is not broken in the
electroweak theory. Therefore, in order to get the gauge invariant
coupling of the superfields, we can simply take the ordinary gauge
invariant electroweak coupling from [30], sect.4.6, and substitute
the ordinary gauge fields $W_1^\mu$, $W_2^\mu, Z^\mu, A^\mu,$ ghost
and scalar fields by the corresponding superfields. Then we obtain the
following triple gauge coupling:
$$T_1^W(x,\theta,\bar\theta)=-{g\over \sqrt{2}}\Bigl[W_\mu^
\dagger W_\nu\Bigl(e^{i\vartheta}V^{\mu\nu}+e^{-i\vartheta}V^
{\mu\nu\dagger}\Bigl)$$
$$ +(W^{\mu\dagger}W_{\nu\mu}-
W^{\mu}W_{\nu\mu}^\dagger)\Bigl(e^{i\vartheta}V^{\nu}
+e^{-i\vartheta}V^{\nu\dagger}\Bigl)\Bigl]
\eqno(4.3)$$
Here $V^{\mu\nu}=\d^\mu V^\nu-\d^\nu V^\mu$ etc. and $\vartheta$ is the
weak mixing angle.

The ghost coupling becomes
$$T_1^U={g\over \sqrt{2}}\Bigl[(W^\mu U_W^\dagger-W^{\mu\dagger}U_W)\Bigl(
e^{i\vartheta}\d_\mu\tilde U_V-e^{-i\vartheta}\d_\mu\tilde U_V^\dagger
\Bigl)$$
$$+\Bigl(e^{i\vartheta}V^\mu+e^{-i\vartheta}V^{\mu\dagger}\Bigl)
(U_W\d_\mu\tilde U_W^\dagger+U_W^\dagger\d_\mu\tilde U_W)$$
$$
+\Bigl(e^{i\vartheta}U_V+e^{-i\vartheta}U_V^\dagger\Bigl)(W^{\mu\dagger}
\d_\mu\tilde U_W+W^{\mu}\d_\mu\tilde U_W^\dagger)\Bigl],
\eqno(4.4)$$
where the ghost and anti-ghost superfields are defined by
$$
U_W=U_1-iU_2,\quad U_V=U_3-iU_4$$
$$
\tilde U_W=\tilde U_1-i\tilde U_2,\quad\tilde U_W^\dagger=-\tilde U_1
-i\tilde U_2.\eqno(4.5)$$
The indices 1,2,3,4 refer to the gauge fields $W_1^\mu, W_2^\mu, Z^\mu$
and the photon $A^\mu$.

In addition to the above couplings we need couplings $T_1^\Phi$ to
unphysical (ghost) scalar fields
$\Phi_1, \Phi_2, \Phi_3$. $\Phi_1$ and $\Phi_2$
form a scalar superfield (3.12)
$$\Phi=(\Phi_1-i\Phi_2)/\sqrt{2},\eqno(4.6)$$
but $\Phi_3$ (the partner of $Z$) becomes a Hermitian superfield
$$\Phi_3=(\Phi_V+\Phi_V^\dagger)/\sqrt{2},\eqno(4.7)$$
because the massless photon has no scalar partner. The scalar couplings
([30], eq.(4.6.9)) then read
$$T_1^\Phi=-{g\over\sqrt{2}}\Bigl\{(\Phi\d_\nu\Phi^\dagger-\Phi^\dagger
\d_\nu\Phi)\Bigl[e^{i\vartheta}V^\nu+e^{-i\vartheta}V^{\nu\dagger}$$
$$-{1\over 2\cos\vartheta}(V^\nu+V^{\nu\dagger})\Bigl]+{1\over\sqrt{2}}
\Phi_3(W_\nu^\dagger\d^\nu\Phi-W_\nu\d^\nu\Phi^\dagger)$$
$$+{1\over\sqrt{2}}(\Phi^\dagger W^\nu-\Phi W^{\nu\dagger})\d_\nu\Phi_3
+m_W\Bigl[e^{i\vartheta}V^\nu+e^{-i\vartheta}V^{\nu\dagger}-{1\over
\cos\vartheta}(V^\nu+V^{\nu\dagger})\Bigl]$$
$$\times(\Phi^\dagger W_\nu-\Phi W_\nu^\dagger)+m_W\Bigl[e^{i\vartheta}
U_V+e^{-i\vartheta}U_V^\dagger-{1\over 2\cos\vartheta}(U_V+
U_V^\dagger)\Bigl]$$
$$\times(\tilde U_W\Phi^\dagger+\tilde U_W^\dagger\Phi)+{m_W\over\sqrt{2}}
(\tilde U_WU_W^\dagger+\tilde U_W^\dagger U_W)\Phi_3+{m_Z\over\sqrt{2}}
\tilde U_3(U_W\Phi^\dagger-U_W^\dagger\Phi)\Bigl\}.\eqno(4.8)$$

All these couplings are determined by the requirement of
quantum gauge invariance, which, in first order means that the total
coupling $T_1=T_1^W+T_1^U+T_1^\Phi$ satisfies
$$[Q, T_1]=i\d_\mu T_1^\mu.
\eqno(4.9)$$
Here, according to (3.52), the (non-vanishing) gauge variations are
given by
$$[Q,W^\mu]=i\d^\mu U_W,\quad [Q,V^\mu]=i\d^\mu U_V$$
$$\{Q,\tilde U_W\}=-i(\d^\mu W_\mu+m_W\Phi),$$
$$\{Q,\tilde U_V\}=-i(\d^\mu V_\mu+m_Z\Phi_3),$$
$$[Q,\Phi]=im_WU_W,\quad [Q,\Phi_3]=im_ZU_3.\eqno(4.10)$$
The so-called $Q$-vertex $T_1^\mu$ in (4.9) can also be taken over from
ordinary gauge theory ([30], eq.(4.3.17)):
$$T_1^\mu={g\over\sqrt{2}}\Bigl\{(W_\nu U_W^\dagger-W_\nu^\dagger U_W)
(e^{i\vartheta}V^{\nu\mu}+e^{-i\vartheta}V^{\nu\mu\dagger})$$
$$+(e^{i\vartheta}U_V+e^{-i\vartheta}U_V^\dagger)(W^{\nu\mu}W_\nu^
\dagger-W^{\nu\mu\dagger}W_\nu)$$
$$+(W^{\nu\mu\dagger}U_W-W^{\nu\mu}U_W^\dagger)(e^{i\vartheta}V_\nu
+e^{-i\vartheta}V_\nu^\dagger)$$
$$-U_W^\dagger U_W\d^\mu(e^{i\vartheta}\tilde U_V-e^{-i\vartheta}
\tilde U_V^\dagger)$$
$$-(e^{i\vartheta}U_V+e^{-i\vartheta}U_V^\dagger)(U_W\d^\mu\tilde
U_W^\dagger+U_W^\dagger\d^\mu\tilde U_W)$$
$$+{1\over 2}(\Phi U_W^\dagger-\Phi^\dagger U_W)\d^\mu(\Phi_V+
\Phi_V^\dagger)+{m_W\over 2}(W^\mu U_W^\dagger-W^{\mu\dagger} U_W)(\Phi_V+
\Phi_V^\dagger)$$
$$+m_W(W^{\mu\dagger}\Phi-W^\mu\Phi^\dagger)\Bigl[e^{i\vartheta}U_V+
e^{-i\vartheta}U_V^\dagger-{1\over 2\cos\vartheta}(U_V+U_V^\dagger)\Bigl]$$
$$+(\Phi^\dagger\d^\mu\Phi-\Phi\d^\mu\Phi^\dagger)\Bigl[e^{i\vartheta}U_V+
e^{-i\vartheta}U_V^\dagger-{1\over 2\cos\vartheta}(U_V+U_V^\dagger)\Bigl]$$
$$+m_W(\Phi U_W^\dagger-\Phi^\dagger U_W)\Bigl[-{1\over\cos\vartheta}
(V+V^\dagger)+i\sin\vartheta (V-V^\dagger)\Bigl]\Bigl\}.\eqno(4.11)$$
Now it is straightforward to check first-order gauge invariance (4.9).
However, second-order gauge invariance requires additional couplings to
physical scalar (Higgs) fields. But the analysis of second-order gauge
invariance is quite different from ordinary gauge theory so that we will
discuss it elsewhere.

The most interesting question is where the supersymmetric extension
differs from the usual electroweak theory. We will find such a
difference if we investigate the self-coupling of the gauge fields (4.3)
in detail. From the first equation in (\ref{RS}) we have the following
expansion in components:
$$
W_\mu=w_\mu+2\bar\theta_{\bar a}\bar\psi_\mu^{\bar a}+i(\theta\sigma
^\nu\bar\theta)\d_\nu w_\mu-m_W(\bar\theta\bar\theta)w_\mu^\dagger$$
$$
-m_W(\bar\theta\bar\theta)\theta^a\psi_{\mu a}+{m_W^2\over 4}
(\theta\theta)(\bar\theta\bar\theta)w_\mu,
\eqno(4.12)$$
and similarly for the adjoint of the second gauge superfield
$$
V_\mu^\dagger=v_\mu^\dagger-2\theta^a\phi_{\mu a}-i(\theta\sigma
^\nu\bar\theta)\d_\nu v_\mu^\dagger-m_Z(\theta\theta)v_\mu$$
$$
-m_Z(\theta\theta)(\bar\theta\phi_{\mu})+{m_Z^2\over 4}
(\theta\theta)(\bar\theta\bar\theta)v_\mu^\dagger.
\eqno(4.13)$$
The interaction Lagrangian in the usual sense is obtained as the
$(\theta\theta)(\bar\theta\bar\theta)$-term in
$T_1(x,\theta,\bar\theta)$, because this is the only term which
contributes if we integrate over $d\theta^2\, d\bar\theta^2$. We substitute
(4.12)-(4.13) into (4.3) and collect the terms with $(\theta\theta), (\bar
\theta\bar\theta)$. The result can be written in the form
$$\int d\theta^2d\bar\theta^2T_1W=-{g\over\sqrt{2}}\Bigl\{{m_Z^2\over 4}\Bigl[
w_\mu^+w_\nu B^{\mu\nu}+w^{\mu\dagger}w_{\nu\mu}B^\nu-w^\mu w_{\nu\mu}
^\dagger B^\nu\Bigl]\eqno(4.14a)$$
$$-{1\over 2}\d_\alpha w_\mu^\dagger w_\nu\d^\alpha B^{\mu\nu}-{1\over
2}(\d_\alpha w^{\mu\dagger}w_{\nu\mu}-\d_\alpha w^\mu w_{\nu\mu}^
{\dagger})\d^\alpha B^\nu $$
$$-{1\over 2}(\d_\alpha w^{\mu\dagger}w_{\nu\mu}+\d_\alpha w^\mu w_{\nu\mu}^
{\dagger}+w^{\mu\dagger}\d_\alpha w_{\nu\mu}+w^\mu\d_\alpha w_{\nu\mu}
^\dagger)\d^\alpha C^\nu$$
$$-{1\over
2}(w_\mu^{\dagger}\d_\alpha w_\nu+w_\mu\d_\alpha w_\nu^
{\dagger})\d^\alpha C^{\mu\nu}\Bigl\}+\ldots .\eqno(4.14b)$$
Here
$$B^\nu=e^{i\vartheta}v^\nu+e^{-i\vartheta}v^{\nu\dagger}=2(\cos\vartheta Z^\nu+\sin\vartheta
A^\nu)\eqno(4.15)$$
is the usual Hermitian combination of the two neutral gauge bosons and
$$C^\nu=e^{i\vartheta}v^\nu-e^{-i\vartheta}v^{\nu\dagger}=2i(\sin\vartheta Z^\nu+\cos\vartheta
A^\nu)\eqno(4.16)$$
is the corresponding imaginary part. The dots in (4.10) represent
contributions coming
from the spin-3/2 components. These couplings are small if those components
have big masses.

The first part (4.14a) of the result has the form of the ordinary
Yang-Mills coupling of the standard model electro-weak theory. The
additional couplings with two derivatives $\d_\alpha$
seems to be anomalous couplings extensively studied in the literature
\cite{HPZH}, 
\cite{DDRW}. However, in our case they can be
transformed by forming divergences and using the Klein-Gordon equation
$\d^2 f=-m^2 f$ in the following way:
$$\d_\alpha f_1\d^\alpha f_2f_3={\rm div}-f_1\d^2 f_2f_3-f_1\d_\alpha
f_2\d^\alpha f_3$$
$$={\rm div}+m_2^2f_1 f_2f_3-f_1\d_\alpha f_2\d^\alpha f_3$$
$$={\rm div}-\d^2f_1 f_2f_3-\d_\alpha f_1f_2\d^\alpha f_3$$
$$={\rm div}+m_1^2f_1 f_2f_3-\d_\alpha f_1f_2\d^\alpha f_3$$
$$={\rm div}+m_1^2f_1 f_2f_3+f_1\d_\alpha f_2\d^\alpha f_3-m_3^2
f_1f_2f_3.$$
Adding the second and the last equation we get the identity
$$\d_\alpha f_1\d^\alpha f_2f_3={1\over 2}(m_1^2+m_2^2-m_3^2)f_1f_2
f_3+{\rm div}.\eqno(4.17)$$
Using $m_W=m_Z\cos\vartheta$, the result (4.14) then assumes the form
$$=-{g\over\sqrt{2}}{m_Z^2\over 2}(1-\cos\vartheta)\Bigl[
w_\mu^+w_\nu B^{\mu\nu}+w^{\mu\dagger}w_{\nu\mu}B^\nu-w^\mu w_{\nu\mu}
^\dagger B^\nu\Bigl]+{\rm div.}\eqno(4.18)$$
Since the divergences lead to a physically equivalent S-matrix, we see
that there are no anomalous couplings. That means, so far our model
agrees with the present experimental data [10] so that it must be taken
seriously.

\end{document}